\documentstyle[aps,multicol,epsfig,eqsecnum]{revtex} 
\input epsf.sty
\input psfig.sty
\def\arQ{Q}
\def\spv{{\bf n}}
\def\Spv{{\bf N}}
\def\pov{{\bf R}}
\def\sov{{\bf g}}

\def\phQ{\phi_{0}}
\def\pcd{q_{\rm P}}
\def\bra#1{\langle#1\vert}
\def\ket#1{\vert#1\rangle}

\def\ipr#1#2{\langle#1\vert#2\rangle}
\def\me#1#2#3{\langle#1\vert#2\vert#3\rangle}   
\newcommand{\eqbreak}{
\end{multicols}
\widetext
\noindent
\rule{.48\linewidth}{.1mm}\rule{.1mm}{.1cm}
}
\newcommand{\eqresume}{
\noindent
\rule{.52\linewidth}{.0mm}\rule[-.1cm]{.1mm}{.1cm}\rule{.48\linewidth}{.1mm}
\begin{multicols}{2}
\narrowtext
}    
\newcommand{\eqbreaknl}{
\end{multicols}
\widetext
\noindent
}
\newcommand{\eqresumenl}{
\noindent
\begin{multicols}{2}
\narrowtext
}    \begin{document}
\draft 

\title{Charge Transport in Manganites: Hopping Conduction, \\ 
the Anomalous Hall Effect and Universal Scaling}
\author{Y. Lyanda-Geller\rlap,\cite{bylineYLG} 
S. H. Chun\rlap,\cite{bylineSHC} 
M. B. Salamon\rlap,\cite{bylineMBS} 
P. M. Goldbart\rlap,\cite{bylinePMG} 
P. D. Han}
\address{Department of Physics and Materials Research Laboratory, \\
University of Illinois at Urbana-Champaign, Urbana, Illinois 61801, USA}
\author{Y. Tomioka, 
A. Asamitsu and 
Y. Tokura}
\address{Joint Research Center for Atomic Technology (JRCAT), 
Tsukuba 305, Japan}
\date{\today}
\maketitle

\begin{abstract}
The low-temperature Hall resistivity $\rho_{xy}$ of
La$_{2/3}$A$_{1/3}$MnO$_3$ single crystals 
(where A stands for Ca, Pb and Ca, or Sr) 
can be separated into Ordinary and Anomalous contributions, 
giving rise to Ordinary and Anomalous Hall effects, respectively. 
However, no such decomposition is possible near the Curie temperature 
which, in these systems, is close to metal-to-insulator transition. 
Rather, for all of these compounds and to a good approximation, 
the $\rho_{xy}$ data at various temperatures and magnetic fields 
collapse (up to an overall scale), on to a single function of 
the reduced magnetization $m\equiv M/M_{sat}$, the extremum of 
this function lying at $m\approx 0.4$. 
A new mechanism for the Anomalous Hall Effect in the inelastic 
hopping regime, which reproduces these scaling curves, is identified. 
This mechanism, which is an extension of Holstein's model for the 
Ordinary Hall effect in the hopping regime, arises from the combined 
effects of the double-exchange-induced quantal phase in triads 
of Mn ions and spin-orbit interactions.  
We identify processes that lead to the Anomalous Hall Effect for 
localized carriers and, along the way, analyze issues of quantum 
interference in the presence of phonon-assisted hopping.  
Our results suggest that, near the ferromagnet-to-paramagnet 
transition, it is appropriate to describe transport in manganites 
in terms of carrier hopping between states that are localized 
due to combined effect of magnetic and non-magnetic disorder.  
We attribute the qualitative variations in resistivity 
characteristics across manganite compounds to the differing 
strengths of their carrier self-trapping, and conclude that both 
disorder-induced localization and self-trapping effects are 
important for transport.
\end{abstract}

\begin{multicols}{2}
\narrowtext
\section{Introduction and overview}
\label{SEC:Intro}
\begin{figure}
\vskip0.7truecm
\epsfig{figure=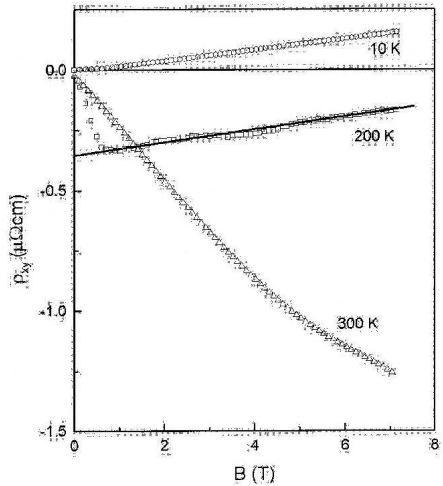,width=2.45in,rheight=2.2in,
angle=0,silent=}    
 \vskip1.4truecm
\caption{Hall resistivity of manganites versus magnetic field 
for a selection of temperatures.  
At $10\,{\rm K}$ the Hall effect is {\it ordinary\/}; the slope 
extrapolates to the origin.
At $200\,{\rm K}$ the Hall effect has both {\it ordinary\/} and 
{\it anomalous\/} components; the slope does not extrapolate to 
the origin, the offset signaling the anomalous Hall effect. 
At $300\,{\rm K}$ it is not simple to separate the Hall resistivity 
into {\it ordinary\/} and {\it anomalous\/} components.}
\label{FIG:introf}
\end{figure}

Numerous recent studies have focused on the Hall effect in the family of 
doped manganese oxides La$_{1-x}$A$_x$MnO$_3$ (in which A stands for Ca, Sr 
or Pb), famous for its colossal magnetoresistance (CMR)~\cite{Jin,Ramirez} 
and accompanying ferromagnet-to-paramagnet (FP) and metal-insulator (MI) 
transitions~\cite
{Snyder,Jaime-PRL,Wagner,Matl,Jakob,Asamitsu,LPMO-PRB,LPMO-PRL,Yuli,LCMOxover}
.  Despite substantial variations in, e.g., the ferromagnet-to-paramagnet 
transition temperature $T_C$ and 
residual resistivity 
across this manganite family, measurements of the Hall effect reveal  
unusual features in both their metallic and insulating regimes. 
An example of the Hall effect data is shown in Fig.~\ref{FIG:introf}.  
In the metallic state the Hall (i.e.~transverse) resistivity $\rho_{xy}$ 
at lowest temperatures (curve at $10\,{\rm K}$ in Fig.~\ref{FIG:introf}) 
exhibit just the ordinary Hall effect (OHE), 
proportional to the external magnetic field $B$. 
At higher temperatures in the 
metallic phase, the Hall resistivity can 
be separated into the sum of: (i)~a (positive) ordinary Hall 
effect, and (ii)~a (negative) 
anomalous Hall effect (AHE), proportional to the magnetization $M$, as 
shown for the curve at 200K on  Fig.~\ref{FIG:introf}. 
The effective density of carrier holes, as deduced from the slope 
of OH resistivity, is typically found to be several times larger 
than that set by the nominal doping level. 

This difference has been 
attributed to the effects of charge compensation and Fermi-surface 
shape~\cite{LPMO-PRB}.  
The AHE is commonly observed in ferromagnets, but
the sign and the magnitude of the AHE in manganites stand in contradiction
to conventional theories based on skew-scattering~\cite{Mott,Smit,Luttinger,Maranzana} 
or side-jump processes~\cite{Luttinger,Berger,Nozieres,YuliZ}. 
Most striking is the rapid increase in the prominence of the AHE that 
occurs at temperatures $T$ close to $T_C$. 
In this range of temperatures, $\rho _{xy}$ can no longer be simply
separated into ordinary and anomalous parts, as can be seen from 
the curve at $300\,{\rm K}$ in Fig.~\ref{FIG:introf}.  For temperatures well 
above $T_C$, $\rho_{xy}$ again becomes linear in $B$, although its 
sign is now negative
~\cite{Jaime-PRL,Wagner,Matl,Jakob,LPMO-PRB,LPMO-PRL,LCMOxover}. 
The corresponding Hall coefficient $R_H$ ($\equiv\rho_{xy}/B$) decreases 
exponentially with increasing temperature in this regime, and a previous study 
identified a clear crossover from non-polaronic to polaronic charge transport 
at around 1.4 $T_C$~\cite{LCMOxover}.

The purpose of the present Paper is to address the issue of charge-carrier 
motion in manganites, both experimentally and theoretically, focusing on 
the vicinity of the FP and MI transitions, from the vantage point 
afforded by the Hall effect.  Our experimental results have led us to 
focus on the anomalous contribution to the Hall effect, 
and to develop a microscopic 
theoretical picture of the charge-carrier motion that gives rise to this 
contribution in manganites.  The picture that emerges is one in which the 
essential character of charge-carrier motion is inelastic hopping between 
states localized due to magnetic and other sources of disorder.
\noindent 

In order to explain the Hall effect in the manganites in the vicinity 
of $T_C$, it is necessary to understand how the nature of the 
charge-carrier states are influenced by the magnetic order of the system.  
In this regard, the double-exchange interaction (DEI), which makes 
charge-carrier motion of Mn outer-shell carriers 
sensitive to magnetic alignment of core 3/2 spins of Mn ions (Hund rules 
lead to alignment of spins in three inner orbitals resulting in 
core spin 3/2), has long 
been known to play a key role in transport in manganites, having been 
introduced by Zener~\cite{Zener} and elaborated by Anderson and 
Hasegawa~\cite{Anderson} and De Gennes~\cite{DeGennes}. 
Therefore, our approach to exploration of the anomalous Hall effect 
in manganites is based on the picture of hopping conduction in the 
presence of double-exchange interaction.

A picture of the ordinary Hall effect in hopping conductors was 
developed long ago by Holstein~\cite{REF:Holstein1}, in which the 
critical ingredient is the Aharonov-Bohm quantal phase~\cite{Aharonov} 
acquired as charge-carriers hop in the presence of magnetic field around 
closed sequences of localized states.  The theoretical work reported 
here amounts to the generalization of Holstein's ideas suited to DE 
systems.  The primary distinctions from Holstein's ideas are as follows:
(i)~Localization is now, to a great extent, caused by magnetic 
disorder in the orientation of core spins (and the attendant 
randomization of hopping amplitudes); the effects of magnetic disorder 
are facilitated by static disorder, and accompanied by polaronic effects. 
(ii)~The relevant quantum-mechanical phases now arise via the quantal 
version of the Pancharatnam phase~\cite{REF:Pancha,REF:BonPan}. 
(iii)~In order for a net Hall effect to result, account must be taken of 
Dzyaloshinskii-Moriya spin-orbit coupling~\cite{REF:Dzyaloshinski}. 
The AHE mechanism that we propose arises in hopping regime in 
systems with localized states, and is the only possible microscopic 
mechanism of AHE in such systems.  A brief account of this work was 
published in Refs.~\cite{LPMO-PRL,Yuli}. 

From the perspective of symmetry, it is well-known that 
spin-orbit interactions lead to 
AHE~\cite{Mott,Smit,Luttinger,Maranzana,Berger,Nozieres,YuliZ}. 
The appreciation that a spin-generated geometric phase, in addition 
to spin-orbit interactions, is an essential ingredient for the 
development of a theory of the AHE in DE systems dates back to the 
1998 manuscript of Kim, Majumdar, Millis and Shraiman~\cite{REF:Kim}. 
From the perspective of the microscopic mechanism of the AHE,
both Ref.~\cite{REF:Kim} and the paper of Ye et al.~\cite{Ye} that 
superceded it, invoke a field-theoretic scheme for integrating 
out the charge-carrier motion and, therefore, were intended for 
metallic regimes (i.e.~regimes in which charge transport occurs via 
{\it delocalized\/} states).  By contrast, the present work 
considers nonmetallic regimes (i.e.~regimes in which charge 
transport occurs via inelastically-assited hopping between 
{\it localized\/} states).  Elsewhere, we shall address the issue 
of the microscopic mechanism of the AHE in the metallic 
regime~\cite{REF:YLG}.   

The microscopic mechanism of AHE that we propose for systems 
with localized states leads to 
remarkable prediction: the Hall resistivity $\rho_{xy}$ depends on the 
temperature and the magnetic field solely through the magnetization $M$, 
i.e., $\rho_{xy}=\rho_{xy}\big(M(H,T)\big)$.  This universal scaling has 
been observed experimentally in manganites. Here, we provide a 
detailed discussion of 
our theoretical picture of hopping transport in manganites.  Further, 
the present paper reports on measurements made on additional compounds 
having lower and higher transition temperatures and provides an analysis 
of these data in terms of our theoretical picture~\cite{LPMO-PRL,Yuli}. 
The universal scaling relation between $\rho_{xy}$ and $M$ reported for 
is shown to hold for the manganese oxides 
${\rm La}_{2/3}{\rm Ca}_{1/3}{\rm MnO}_{3}$ (LCMO), 
${\rm La}_{2/3}{\rm Sr}_{1/3}{\rm MnO}_{3}$ (LSMO), and 
${\rm La}_{2/3}{\rm Ca}_{1/9}{\rm Pb}_{2/9}{\rm MnO}_{3}$ (LPMO). 
Although data on the Hall effect in these compounds have universal features,
the the temperature dependence of resistivity in 
LSMO is different from that in 
LCMO and LPMO. 
This behavior is due to different size of dopant ions which results in 
different static disorder, different carrier localization length, 
and, accordingly, different strength of self-trapping due to lattice 
effects. We believe that the accuracy of the results concerning the 
AHE based on inelastic charge-carrier hopping between states localized 
due to magnetic and nonmagnetic sources of disorder suggests that the 
dominant mechanism for charge transport in the transition regime is 
indeed inelastic charge-carrier hopping between localized states, 
which differs qualitatively from the picture of metallic conduction 
perturbed by double-exchange interactions. 
As for polaronic effects, depending on the compound, they may set in 
soon as localization length is of the order 
of lattice constant. These effects (or their absence) are crucial 
for the character of the temperature behavior in the range of 
high temperatures above the FP and MI transitions. 
Polaronic effects do not affect the universal scaling 
of the Anomalous Hall resistivity. At the same time, scaling of the 
AH resistivity of the type observed in the CMR regime is not contained 
in conventional 
models of the AHE in the metals (i.e., those based on skew-scattering 
and side-jump mechanisms).  Neither is this scaling contained in a Berry 
phase mechanism for the AHE in the metallic phase, discussed in 
Refs.~\cite{REF:Kim,Ye}. We regard this as further evidence against the 
viability of any metallic-based picture of transport in the transition regime.

How does the present work relate to earlier work on charge transport in 
the CMR regime?  Attempting to build on the early key insight that DE 
plays a central role, Millis et al.~\cite{Millis1} considered transport 
in DE systems within the framework of the coherent potential approximation 
(CPA).  Making the CPA in the present context amounts to replacing the 
charge and magnetic system by an effective one, involving only the charge 
subsystem, in which the conduction band width depends on the magnetization 
but there is no other effect of the magnetic sector.  Thus, any resistivity 
obtained via such a scheme is simply whatever the resistivity of the charge 
sector was, reduced by an extent that depends on the magnetic order 
via a renormalization of the bandwidth.  The picture enforced by this 
approach is that the fundamental mechanism for charge transport is 
metallic in nature.  Naturally, the CPA approach~\cite{Millis1} is 
unable to yield a colossal magnetoresistance, although it can provide 
factors on the order of unity.  What it explicitly omits is any 
resistivity mechanism arising from localization due to magnetic disorder, 
as noted by Varma~\cite{Varma}.  Rather than appeal to such a localization 
process, Millis et al.~\cite{Millis2} proposed that the 
magnetization-dependent reduction of the bandwidth invites lattice 
effects.  Specifically, the (now magnetically) heavy 
charge carriers would be more susceptible to self-trapping by a large 
Jahn-Teller lattice distortion, which would cause a metal-insulator 
transition via polaronic collapse of the conduction bandwidth. 

Accepting, for the moment, the notion that charge transport in the 
transition regime is indeed accomplished by lattice polarons, 
let us ask what the consequences would be for the resistivity. 
According to theory of polaronic transport, developed in series of 
papers in 1960's by Holstein alone~\cite{REF:Holstein}, with 
Friedman~\cite{REF:Friedman} and with Emin~\cite{REF:Emin}, polaronic-type 
conduction manifests itself via a specific temperature-dependence 
of the longitudinal and Hall resistivities, being activated in 
character with a definite relationship between the activation 
constants for these resistivities.  Following the proposal of 
polaronic-type conduction by Millis et al., experimental tests of 
these temperature dependences were performed.  
Initial results~\cite{Jaime-PRL} in LPMO  
over the range of temperatures high 
above the MI and FP transitions 
seemed in accordance with the polaronic picture.  However, recent 
extensive measurements at lower temperatures, 
in transition regime~\cite{LCMOxover}, demonstrate 
that, at least in this regime, the temperature dependence of the 
longitudinal and Hall resistivities cannot be explained in terms of  
polaronic picture alone. Furthermore,
even at high temperatures, 
polaron-based picture is not compatible with experimental data 
for LSMO samples.

With the pictures of charge transport in the CMR regime based on 
either the magnetization-dependent reduction of the bandwidth or  
on polarons alone invalidated, what remains is the possibility of constructing 
a valid picture based on non-polaronic localization of charge carriers.  
Strong evidence supporting such a picture comes from a simple estimate 
of the scattering time (i.e.~the scattering-induced conduction-band 
broadening), which indicates that, in the transition regime, the band 
broadening exceeds the band width (i.e.~the mean free path is shorter 
than the Fermi wavelength) so that the resistivity exceeds the 
Mott-Ioffe-Regel limit and, hence, the conduction cannot be metallic.
Therefore, one needs to search for insulating transport mechanisms 
and, specifically, the origins of carrier localization that are distinct 
from polaronic effects. (We note that in compounds in which the Jahn-Teller 
distortion is not symmetry-allowed, this is especially important). Such 
localization can result from both magnetic disorder (i.e.~due to lack 
of core-spin alignment) and non-magnetic disorder (i.e.~static potential 
disorder due, e.g., to doping)~\cite{Sheng}.  

While the localizing influence of non-magnetic disorder on charge 
transport has been thoroughly investigated~\cite{REF:Efros,REF:LeeRama}, 
the influence of magnetic disorder is less well known, and we shall 
discuss it in detail in Sec.~\ref{SEC:Dis_loc}.  For now, we simply mention 
that the magnetic disorder in core 3/2 spin orientation 
experienced by the outer-shell charge 
carriers arises via the DE interaction from fluctuations around the 
ferromagnetic state that build up as the FP transition is approached from 
the low temperature side.  Of course, these fluctuations are 
dynamical, but they are slow, compared with characteristic 
timescales for outer-shell charge-carrier motion.  Thus, for the purposes 
of analyzing the influence of the magnetic sector on charge transport, 
it is appropriate to regard orientations of the core spins on the Mn ions 
as quenched variables.  The resulting magnetic disorder takes the form of 
randomness in the off-diagonal hopping matrix elements for the charge 
carriers. By contrast, nonmagnetic disorder occurs due to randomness 
in the substitution of La by dopant ions (e.g.~Sr, Pb or Ca), and gives 
rise to the more familiar diagonal (Anderson-type) disorder. 
Electronic states in systems with off-diagonal disorder 
were first considered by Lifshitz~\cite{REF:Lifshitz}, who showed that  
localized states arise in the band tail.  The physical picture of carrier 
states in manganites must encompass both magnetic and nonmagnetic disorder,
possibly facilitated by Coulomb effects, which (jointly or severally) can 
result in carrier localization. 

If carriers are localized then they can still participate in transport, 
but it is by hopping from one localized state to another, assisted by 
one or more inelastic agents (such as phonons).  In this case, the 
longitudinal resistivity is determined by the rate of inelastic 
hopping between occupied and unoccupied states~\cite{REF:Efros,REF:AHL}. 
When carrier localization has occurred, and the localization length 
is of order of lattice constant, electronic interaction with lattice 
and self-trapping effects can become essential,
so that at high temperature resistivity is determined by small polarons. 
However, transitional regime is greatly affected by carrier 
localization of non-polaronic origin. We notice that in general one 
should distinguish between Jahn-Teller polarons and Holstein breathing mode 
polarons: the presence of the former depends on symmetry of the system, 
the latter arise independent of the underlying symmetry. In manganites,
both types of polarons are capable of facilitating carrier localization 
by magnetic and non-magnetic disorder; when carriers are localized on 
lattice constant scale, Holstein polarons govern the temperature 
dependence of resistivity deep in the insulating phase.

In the present Paper, we shall not consider metallic manganites, and 
restrict our consideration to the 
inelastic hopping transport regime. The discussion of the Hall effect in 
metallic ferromagnets will be presented elsewhere~\cite{REF:YLG}. 
The present Paper is organized as follows. 
In Sec.~\ref{SEC:exp} we describe the experimental setup and in 
Sec.~\ref{SEC:expres} we present experimental data on the longitudinal 
resistivity, magnetization and  Hall resistivity in three different 
manganite compounds having distinct transition temperatures.  
Section~\ref{SEC:theory} is organized into several subsections, in which 
we describe models of disorder in manganites, issues related to the 
localization of carriers and the hopping transport mechanism, as well 
as the quantal Pancharatnam phase, spin-orbit Dzyaloshinski-Moriya 
interactions, and the universal scaling of the Hall resistivity. 
In Sec.~\ref{SEC:disctheorex} we discuss the correspondence between our 
theoretical and experimental results.  

\section{Experiments on transport and magnetic properties of manganites}
\label{SEC:exp}

\subsection{Experimental method}
\label{SEC:expmethod}

In the experimental part of this study, single crystals of various 
manganites were used in order to avoid extrinsic effects from grain 
boundaries or strains. In single crystals, simultaneous measurements 
of transport and magnetic properties permits us to find a precise 
dependence of transport coefficients on the sample magnetization.  
We have measured the 
longitudinal and Hall resistivities and the magnetization of three 
different crystals with various transition temperatures. 
La$_{0.7}$Ca$_{0.3}$MnO$_3$ (LCMO) and 
La$_{0.7}$Sr$_{0.3}$MnO$_3$ (LSMO) single crystals were prepared
by the floating-zone method. 
La$_{0.67}$(Ca,Pb)$_{0.33}$MnO$_3$ (LPMO)
single crystals were grown from 50/50 PbF$_2$/PbO flux. More details 
on the sample growth and basic properties can 
be found 
elsewhere~\cite{LCMOsample,LPMOsample}.  All specimens used in the 
measurements were cut
along crystalline axes into bar shapes from larger pre-oriented crystals.
Contact pads for Hall resistivity measurements were made by sputtering 
$\approx $1500~\AA\ of gold through a mask.   Gold wires of 50 $\mu $m
diameter are then attached using slowly drying silver paints.  Typical
contact resistances after annealing were about 1 $\Omega $ at room
temperature.  We adopted a low-frequency (39~Hz) ac method for the
measurements.   The transverse voltage signal was first nulled at zero 
field at each temperature below 400~K by a potentiometer, and the change 
in the transverse voltage was recorded as $H$ was swept from +7 T to -7~T 
and back for averaging.  Following the transport measurements, sample 
magnetizations were measured by a 7~T SQUID magnetometer on the same 
samples.

\subsection{Experimental Results}
\label{SEC:expres}

\begin{figure}
\epsfig{figure=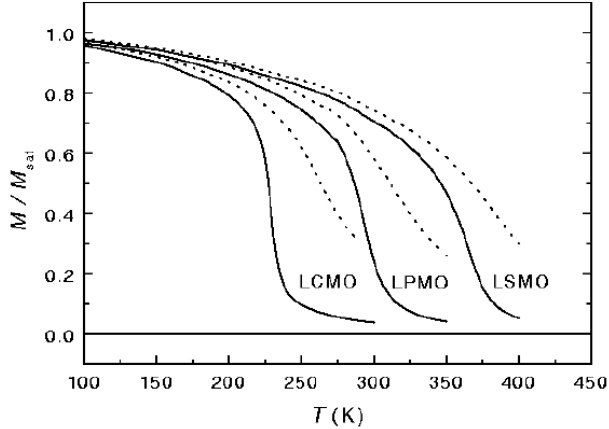,width=3.2in,rheight=3.2in,
angle=-90,silent=}    
\vskip-2.2cm
\caption{Temperature dependences of normalized magnetization 
$M/M_{\rm sat}$ 
under $1\,{\rm T}$ (solid lines) and 
under $7\,{\rm T}$ (dotted lines) magnetic fields.}
\label{FIG:mag}
\end{figure}

Figure~\ref{FIG:mag} shows the temperature dependences of 
magnetization measured at 1~T
and 7~T. All three samples show ferromagnetic-to-paramagnetic phase 
transitions.  The Curie temperatures $T_C$ were determined by scaling 
analysis on high field $M(H)$ curves near the transition, and the results 
are shown in Table~I. As $T_C$ decreases, the transition becomes sharper, 
resulting in anomalous critical exponents~\cite{scaling}. 

\begin{figure}
\vskip-0.5cm
\epsfig{figure=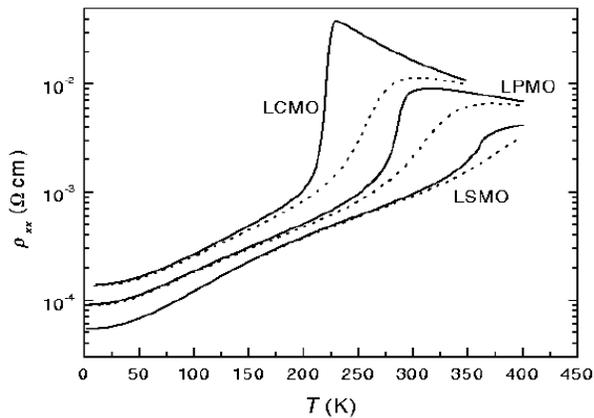,width=3.1in,rheight=3.2in,
angle=-90,silent=}    
\vskip-1.7cm
\caption{Temperature dependences of longitudinal resistivity 
$\rho_{xx}$ under zero magnetic field (solid lines) 
and under $7\,{\rm T}$ (dotted lines).}
\label{FIG:longrho}
\end{figure}

The temperature dependences of the longitudinal resistivities $\rho_{xx}$
for the same set of samples under zero magnetic field and under 7~T are
shown in Fig.~\ref{FIG:longrho}.  
LCMO and LPMO show metal-insulator transitions near $T_C$,
whereas LSMO shows an inflection at $T_C$, but $\rho_{xx}$ 
continues to increase
with increasing temperature above $T_C$. 
The metal-insulator transition 
temperatures $T_{MI}$, determined by the maximum in the rate of change in 
the temperature dependence of the longitudinal resistivity 
$d\rho _{xx}/dT$
under zero magnetic field, were slightly higher than corresponding $T_C$'s
(Table~I).  Also shown in Table~I are $\rho _{xx}$ minima (occurring at 
the lowest temperatures) and magnetoresistivity (MR) maxima 
[defined by $\big(\rho _{xx}(0~{\rm T})-\rho _{xx}(7~{\rm T})\big)/
\rho _{xx}(7~{\rm T})$].  The observed decrease in $T_C$ correlates 
with the overall increases of resistivity and MR, which can be clearly 
seen in Fig.~\ref{FIG:longrho}.

\begin{figure}
\vskip-1.5cm
\epsfig{figure=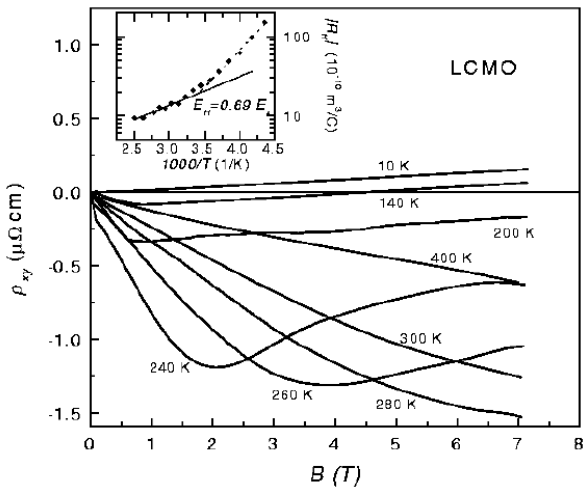,width=3.1in,rheight=3.1in,
angle=-90,silent=}    
\vskip-1.0cm
\caption{Main panel: 
Hall resistivity $\rho_{xy}$ of LCMO as a function of
magnetic field at the indicated temperatures. 
Inset: Activated behavior of the high temperature Hall 
coefficient $R_{H}$.}
\label{FIG:hall1}
\end{figure}

\begin{figure}
\vskip0.5cm
\epsfig{figure=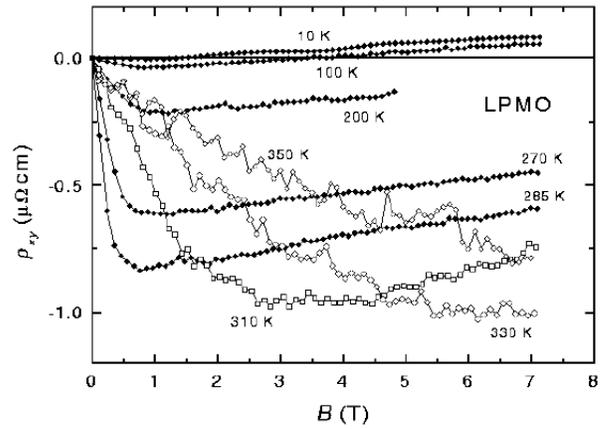,width=3.1in,rheight=3.5in,
angle=-90,silent=}    
\vskip-3.0cm
\caption{Hall resistivity $\rho_{xy}$ of LPMO as a function of
magnetic field at the indicated temperatures.}
\label{FIG:hall2}
\end{figure}

\begin{figure}
\vskip0.5cm
\epsfig{figure=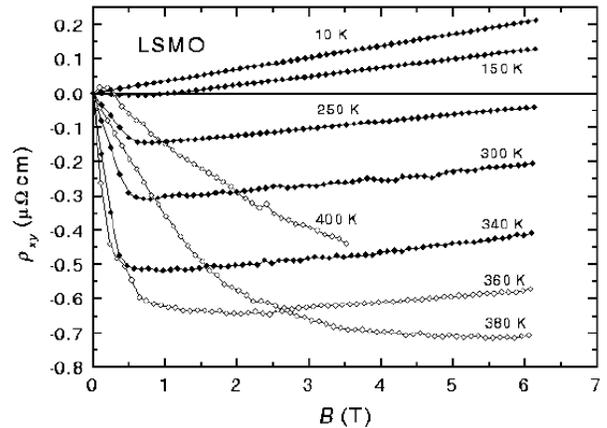,width=3.1in,rheight=3.5in,
angle=-90,silent=}    
\vskip-3.0cm
\caption{Hall resistivity $\rho_{xy}$ of LSMO as a function of
magnetic field at the indicated temperatures.}
\label{FIG:hall3}
\end{figure}

Despite differences in $T_{C}$, $T_{MI}$ and temperature dependence 
of longitudinal resistivity across the three compounds, 
the Hall resistivity of 
compounds with doping that corresponds to maximal $T_{C}$  
show similar temperature and field dependences, as shown in
Figs.~\ref{FIG:hall1}, \ref{FIG:hall2} and \ref{FIG:hall3}. 
At low temperatures, $\rho _{xy}$ is positive and linear in 
magnetic field, the sign indicating hole-like charge carriers, and negligible
anomalous Hall contribution.
\eqbreak

\begin{table}[tbp]
\caption{Characteristics of single crystal samples used in this study}
\label{table1}
\begin{tabular}{llccrrc}
& composition & $T_C$ & $T_{MI}$ & min $\rho _{xx}$ & max MR & $n_{eff}$(10 K) \\ 
\tableline LCMO & La$_{0.7}$Ca$_{0.3}$MnO$_3$ & 216.2 K & 
222.5 K & 140 $\mu \Omega $cm & 2,600 \% & 1.6 holes/Mn \\ 
LPMO & La$_{0.67}$(Ca,Pb)$_{0.33}$MnO$_3$ & 285.1 K & 287.5 K & 91 $ \mu \Omega $cm & 326 \% & 2.4 holes/Mn \\ 
LSMO & La$_{0.7}$Sr$_{0.3}$MnO$_3$ & 359.1 K & 362.0 K & 55 $ \mu \Omega $cm & 64 \% & 1.0 holes/Mn 
\end{tabular}
\end{table}%

\eqresume

\noindent
As the temperature is increased, the high-field
slope is roughly the same.  However, the increasing, negative, contribution 
to $\rho _{xy}$ shifts it downward.  Quantitatively, $\rho _{xy}$ can be 
expressed as a sum of 
an ordinary contribution 
parametrized by $R_0\left(T\right)$ 
and an anomalous contribution parametrized by 
$R_S\left( T\right)$~\cite{Hurd}: 
\begin{equation}
\rho_{xy}\left(B,T\right)=
R_{0}\left(T\right)\,B+
\mu_{0} R_{S}\left(T\right)\,M\left(H,T\right),
\label{eq1}
\end{equation}
where $B=\mu_0 [H+(1-N)M]$, $H$ is the external magnetic field, 
and $N\approx 1$ is the demagnetization factor. 
As has commonly been observed in manganite crystals and thin films, 
the effective charge-carrier density 
$n_{eff}\equiv 1/eR_{0}$ is scattered between 1.0 and 2.4 
holes/Mn (see Table~I), which is much larger than the nominal 
doping level (of 0.3-0.33 holes/Mn), presumably due to the 
effects of the anisotropy of the Fermi surface~\cite{LPMO-PRB}.  
We refer to our previous publications for the discussion on 
the low temperature OHE~\cite{LPMO-PRB}.

On further increase of the temperature through $T_C$, $\rho _{xy}$ 
becomes much larger, strongly curving with magnetic field, and the 
positive, high-magnetic-field contribution, linear in the field,
which would arise from the ordinary Hall effect in a metallic phase due 
to the Lorentz force acting on charge-carriers, disappears.  Owing to 
its low $T_{C}$, for the LCMO sample we were able to explore temperatures 
far above $T_C$, where $\rho_{xy}$ shows a negative Hall coefficient, 
despite the doping of the material being by holes.  In this range of 
temperatures the Hall coefficient $R_{H}=\rho _{xy}(B)/B$ exhibits
activated behavior, with a characteristic energy 
$E_{H}\approx\frac{2}{3}E_{\sigma }$, where $E_{\sigma }$ is the 
activation energy for ordinary conductivity $\sigma_{xx}$.  Similar 
experimental results have been obtained in Refs.~\cite{Jaime-PRL,LCMOxover}. 
In these works, investigations of manganite carrier-transport deep in the 
insulating phase have shown that the sign and temperature dependence of the 
high-temperature (i.e.~above 1.4$T_C$) Hall coefficient $R_H$ 
($\equiv\rho_{xy}/B$) can 
be explained in terms of the adiabatic hopping of small polarons.  

\begin{figure}
\vskip0.5cm
\epsfig{figure=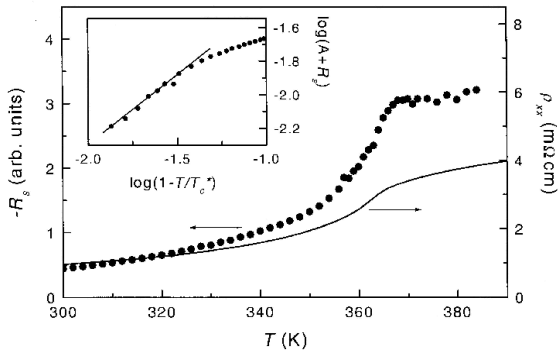,width=3.1in,rheight=3.5in,
angle=-90,silent=}    
 \vskip-3.6cm
\caption{Main panel: 
The anomalous Hall coefficient $R_S$ (symbols) compared 
with the longitudinal resistivity $\rho_{xx}$ (solid line). 
Inset: The critical behavior of $R_S$.}
\label{FIG:check}
\end{figure}

Initially~\cite{Jaime-PRL}, high-temperature transport 
picture due polaron hopping to was believed to support the proposal 
by Millis et al.~\cite{Millis1} that the Jahn-Teller 
distortion which, according to symmetry considerations can occur in the
MnO$_6$ octahedra of LaMnO$_3$, is responsible for the insulating 
behavior of doped La$_{1-x}$A$_x$MnO$_3$ systems. Hall resistivity 
measurements should be capable of providing key evidence for or against 
the polaronic picture of charge transport.   According to the theory of 
this picture, the adiabatic hopping of small polarons~\cite{REF:Emin} leads 
to an activation energy $E_H$ characterizing $R_H$ that is 2/3 of the 
activation energy $E_\rho $ characterizing $\rho _{xx}$~\cite{REF:Emin}, as 
is observed at high temperatures~\cite{Jaime-PRL,LCMOxover}.  However, 
recent Hall resistivity measurements, extending into the transition 
region~\cite{LCMOxover} show that the activation energy changes abruptly 
at a cross-over temperature $1.4T_{C}$, 
from the polaronic value of  $\frac{2}{3}E_{\sigma }$ to a 
much larger value, $1.7E_{\sigma}$.  This clearly marks the breakdown of 
the small polaron picture of charge transport, as shown in 
Fig.~\ref{FIG:hall1} (inset).
In fact, the effective activation energy of the conductivity begins to 
decrease from a value of $E_{\sigma}$ at roughly the same 
cross-over temperature, 
making even greater the discrepancy between the experimental 
data and the small-polaron hopping picture.  Even more dramatically, the 
product of the Hall mobility $\mu_{H}$ and the temperature, viz., 
$\mu_{H}T=-\sigma _{xx}R_{H}T$ which, according to the small polaron 
picture, should decrease monotonically with decreasing temperature, in 
fact is found to exhibit a minimum at the same cross-over temperature. 
(We shall later show that, near $T_C$, $\rho_{xy}$ is determined solely 
by the sample magnetization in all three compounds.)\thinspace\  Thus, 
experiments in transition region lead to the conclusion that, while 
small polarons are an essential part of the physics of transport in 
manganites at high temperatures, they cannot provide a complete picture 
of the metal-to-insulator transition.  
\vskip1.5cm
\vskip-0.3cm
More generally, as discussed by 
Varma~\cite{Varma}, there exist double-exchange systems, such as 
TmSe$_x$Te$_{1-x}$, in which transport phenomena observed in manganese 
oxides are also observed but Jahn-Teller distortions, leading to small 
polarons, are not symmetry-allowed. At the same time, if the carrier 
localization length becomes of order of lattice constant, lattice effects 
in the form of Holstein breathing mode polarons arise naturally. 
This allows to explain why the high-temperature regime in some of 
manganese compounds exhibits 
longitudinal and Hall resistivities characterized by thermally-activated 
behavior which is
qualitatively and quantitatively consistent with that caused by polaronic 
transport mechanism. 
Before turning to the theoretical picture of transport in manganites 
and, specifically, of AHE for localized carriers, we 
pause to examine whether the 
theoretical model of the AHE proposed by Ye et al.~\cite{Ye} which is 
based on a metallic view of charge transport, is consistent with our 
experimental data.  In that model, following 
an earlier model by Kim et al~\cite{REF:Kim}, it is assumed that 
tight-binding charge-carriers propagate coherently through a smoothly 
varying magnetization texture which has the effect of introducing a 
Berry phase gauge 
potential~\cite{REF:LeeNaga,REF:Wiegmann,REF:Dagotto,REF:LAG,footnote}.  
A central prediction of the model due to Ye et al.~is that 
a peak should occur in $R_{S}(T)$ above $T_{C}$, along with 
a singularity in the slope at $T_{C}$, i.e., 
$dR_{S}/dT\sim\vert 1-T/T_{C}\vert^{-\alpha }+c$, 
where $\alpha$ is the specific heat exponent. 
To test this prediction, we measured the low magnetic field 
($<0.5\,{\rm T}$) magnetization and Hall resistivity of our most 
metallic sample, LSMO near $T_{C}$ 
(see Fig.~\ref{FIG:check}). 
From the behavior of $\rho_{xy}$ and $M$ in the zero-field 
limit, we determined $R_S\equiv (d\rho_{xy}/dM)$. 
In constrast to the prior 
report by Matl et al.~\cite{Matl}, in this \lq\lq metal-to-metal\rq\rq\  
transition system we do not find $R_{S}$ to be proportional to $\rho_{xx}$. 
As seen on Fig.~\ref{FIG:check}, $\rho_{xx}$ flattens at the 
temperature at which the resistive transition is complete 
(i.e.~$T_{C}^{*}=368\,{\ rm K}$).  This 
allows us to conclude that neither a constant $R_{S}/\rho_{xx}$ nor a peak 
in $R_{S}$ is a common feature in manganites. We note that 
close to $T_{C}^{*}$, it is possible to express $R_{S}(T)$ as a power law,
$(1-T/T_C^{*})^{0.82}+A$ (see Fig.~\ref{FIG:check}, inset). However above 
$T_{C}^{*}=368 {\rm K}$, which is significantly higher than both $T_{C}$ 
and $T_{MI}$ (see Table I), $R_{S}$ is constant, and the fit does 
not correspond to the inflection point predicted in Ref.~\cite{Ye}.

\section{Theory of hopping magnetotransport in manganites}
\label{SEC:theory}

The aim of the present section is to develop a picture of the Hall effect 
in manganites, to test this picture through comparison with experimental 
data and, hence, to build as completely as possible a general picture of the 
of charge transport in manganites in the 
ferromagnet-to-paramagnet transition regime. 
Among the results we shall obtain, perhaps the most striking is 
the universal scaling of the magnetization-dependent Hall resistivity, 
which we explain should hold in the regime where charge transport 
proceeds primarily via hopping between localized states.  Such 
universal scaling has been observed experimentally.

\subsection{Disorder and interactions in manganites}
\label{SEC:Dis_int}

Manganites are extremely complicated materials, and a bewildering variety 
of behaviors occurs in them, as the doping level, temperature or magnetic field 
is varied.  Here, we focus on those manganites that exhibit a transition 
from a ferromagnetic metal to a paramagnetic insulator, controlled by 
temperature, as occurs in manganite compounds, doped with 
Ca or Sr or (Pb and Ca)  substituting for La, 
at doping levels of around 1/3.  This doping, of 
course, results in several sources of static disorder:  
 (i)~the dopant ions subsitute randomly for La; and 
(ii)~the lattice distortion around the two ionizations of Mn 
(viz.~Mn$^{3+}$ and Mn$^{4+}$) is distinct (i.e.~the breathing-mode 
effect).  This disorder leads to local variations in the amplitudes 
for the hopping processes that carry charges between magnetic ions. 
Furthermore, (iii)~any clustering of dopants in the randomly doped 
lattice would lead to fluctuations in the carrier density.  These 
sources produce {\it nonmagnetic\/} disorder. 

Along with the motion of charge carriers there is also the motion of 
core spins.  In the present context, we believe that it is 
profitable to treat theses core spins classically, and to regard 
the carrier dynamics as being much faster than the spin dynamics, 
so that the carrier motion can be pictured as taking place within a 
frozen core-spin configuration that is randomized owing to  
thermal fluctuations (i.e.~we adopt a quasi-static approach). 
We term such random magnetic configurations \lq\lq magnetic 
disorder\rlap.\rq\rq\thinspace\  We note that spin-spin correlation 
times have been obtained experimantally from muon spin relaxation 
and neutron spin echo data~\cite{Heffner}; these experiments show 
that, indeed, spin dynamics is slow.

Strong thermal fluctuations 
render typical instantaneous configurations of the spins rather 
inhomogeneous.  Among these fluctuations, there are the 
\lq\lq hedgehog\rq\rq\ excitations which, owing to their topological 
stability, are long-lived, and become more numerous as the 
ferromagnet-to-paramagnet transition is approached~\cite{Lau}.  Due 
to the resulting magnetic inhomogeneity, the carrier-hopping matrix 
elements are reduced.  

As we shall discuss in the following subsection, the presence of 
nonmagnetic and magnetic disorder both support the notion that the carrier 
states are localized at temperatures near to the 
(zero-magnetic-field) FP transition, as well as at higher temperatures.  
Such localization of carriers can explain the resistive transition which, 
in turn, leads to the disappearance of double-exchange-induced ferromagnetic 
spin-correlations, at least on spatial scales larger than the localization 
length.  We note that in a series of papers~\cite{Furukawa} Furukawa has 
considered the issues of carrier states and transport in manganites by using 
a dynamical generalization of the coherent potential approximation, arriving 
at the conclusion that the transport properties of manganites can be explained 
in terms of the scattering of metallic carriers by magnetic randomness.  We, 
however, believe that the fact that (in the range of temperatures marking the 
transition regime) the resistivity exceeds the Mott-Ioffe-Regel limit renders 
any approach founded on the scattering of delocalized carrier-states to be  inconsistent. 

As discussed above in Sec.~\ref{SEC:Intro}, localization effects related to 
Jahn-Teller distortions and small-polaron formation cannot explain certain 
central experimental data in the transition regime, including the 
temperature-dependence of both the resistivity and the Hall effect.  
Therefore, one is forced to consider alternative mechanisms that can 
lead to the breakdown of metallic conductivity and can also serve as an 
origin of various universal transport properties that have been observed in double-exchange systems, including, e.g., those in which Jahn-Teller 
distortions are symmetry-forbidden.  For these reasons, we now give a  
discussion of the physics of disorder-induced carrier localization in 
manganites.

\subsection{Disorder-induced carrier localization in manganites}
\label{SEC:Dis_loc}

\begin{figure}[hbt]
\epsfxsize=3in
 \centerline{\epsfbox{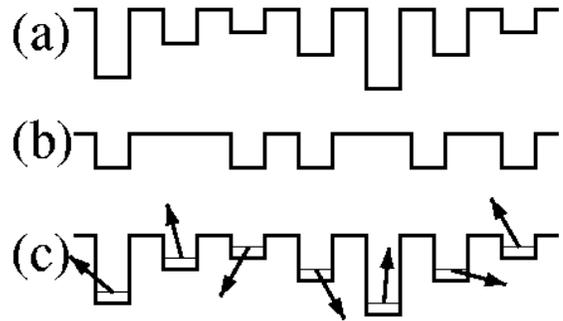}}
\vskip+0.20truecm
\caption{Schematic one-dimensional picture of 
disorder in manganites. 
(a)~Anderson model with on-site disorder; 
(b)~Lifshitz model with random ionic locations 
(leading to random transfer integrals);
(c)~model of disorder in manganites, showing 
on-site disorder and magnetic disorder 
(leading to random transfer integrals)}
\label{FIG:dis}
\end{figure}

To see why it is useful 
to regard charge transport as taking place in a frozen random background 
of core-spin orientations, let us imagine the spin configuration to be 
truly static.  In the transition regime, a typical spin configuration is 
rather inhomogeneous and, hence, we expect carriers to be localized.  
Support for this notion comes from the close similarity between transport 
in manganites and systems of randomly located identical impurities 
(i.e.~off-diagonal disorder).  For the latter, localization has been 
established via the work of Lifshitz~\cite{REF:Lifshitz}.
Although 
spin-induced randomness in manganites [arising from the random 
double-exchange factors of $\cos(\theta/2)$, where $\theta$ is the 
angle between core spins on Mn ions, see, e.g., Sec.~\ref{SEC:qp}] 
is weaker than the randomness considered in Ref.~\cite{REF:Lifshitz}, we 
expect the two systems to exhibit qualitatively similar localization 
behavior.  Furthermore, the condition for localization in the Lifshitz 
model (viz.~that the characteristic spatial scale of the outer-shell wavefunctions in isolated Mn ions be much smaller than distance between 
sites) is well obeyed in manganites.  Therefore, provided that there is 
appreciable randomness in the core-spin orientations, transport 
properties should be determined by the short-distance physics of 
clusters of ions and by magnetic correlations between such clusters.  
Moreover, nonmagnetic disorder and possible states bound to the 
subsituting A-site ions are capable of amplifying the trend towards
localization~\cite{Varma,Sheng}.  In Fig.~\ref{FIG:dis} we present 
a one-dimentional caricature of disorder 
in manganites, in which diagonal and off-diagonal disorder coexist.  
Sheng et al. included both magnetic and non-magnetic disorder and 
applied one-parameter scaling theory~\cite{REF:GangOfFour} and finite-size 
scaling ideas~\cite{Kramer} in order to investigate carrier localization 
in manganites numerically.  Sheng et al.~found that, in the presence of 
magnetic disorder, an Anderson metal-insulator transition accompanies the 
ferromagnet-to-paramagnet transition.  They also observed an interesting 
correlation between $T_C$ and the residual resistivity, which is determined 
by nonmagnetic disorder, viz., the larger the residual resistivity, the 
smaller the $T_C$; this agrees well with the original double-exchange 
picture, in which carrier motion promotes ferromagnetism whereas disorder 
resists electronic motion and, therefore, does not promote ferromagnetism.  

To what extent can one regard charge transport as taking place in a frozen 
spin configuration?  Equivalently, what are the characteristic time 
scales for magnetic and charge dynamics?  For charge dynamics the time 
scale is $\hbar/t$, where $t$ is a characteristic magnitude of the 
hopping matrix element, i.e., the shortest relevant time scale.  For magnetic 
dynamics, the issue is more complicated.  However, the shortest time scale 
is presumably $\hbar/k_{\rm B}T$, which, in the regime of interest to us, 
is greater than $\hbar/t$.  Furthermore, as discussed by Lau and 
Dasgupta~\cite{Lau}, in three-dimensional magnetic systems the 
ferromagnet-to-paramagnet transition involves very long-lived topological 
excitations (i.e.~magnetic hedgehogs)~\cite{Murthy} which, as 
we shall see, are particularly significant for the AHE. 
  Hence, we see that, to a first 
approximation, one can regard charge transport as taking place in a frozen 
spin configuration.  Corrections to charge dynamics, due to magnetic 
dynamics (as well as feedback into the magnetic dynamics sector) can, if 
necessary, be treated by going beyond the Born-Oppenheimer approximation. 

\subsection{Percolation-hopping scenario of 
transport phenomena in manganites}
\label{SEC:Perc_scen}

As discussed in the previous subsection, carrier states in 
manganites are effectively localized throughout the transition 
regime.  (By effectively localized we mean localized on timescales 
short compared with that required for the reconfiguration of the 
magnetic degrees of freedom.)\thinspace\  Therefore, in this regime 
transport of carriers occurs via inelastic hopping (i.e.~hopping 
that is assisted by some inelastic agent such as a phonon).  In the 
following Sec.~\ref{SEC:hop_pic}, we describe a picture of hopping 
transport applied to the setting of manganites.  Following this, in 
Sec~\ref{SEC:hopCP}, we use this picture to discuss a scenario for 
transport phenomena in LCMO and LPMO, which are materials in which 
polaronic effects are believed to be important.  Then, in 
Sec.~\ref{SEC:hopS}, we describe the scenario of hopping transport 
suitable for application to LSMO, a material in which signatures 
of polaronic effects are absent.  By contrasting LCMO and LPMO with 
LSMO we draw some general conclusions concerning the role played by 
polarons in various manganite compounds.

\subsubsection{Hopping transport picture in manganites}
\label{SEC:hop_pic} 

Hopping transport models based on percolation theory have been 
successfully applied to transport in systems with states localized 
by static disorder~\cite{REF:Efros,REF:AHL}.  A peculiarity of 
manganite systems is that the wavefunctions of states localized in 
the vicinity of Mn ions depend on the orientations of the Mn core spins 
on these ions.  Inelastic agents (e.g.~phonons) lead to hopping between 
these localized states, the amplitude for such hopping being 
characterized by matrix elements of the carrier-phonon interaction.  
Therefore, hopping probabilities and rates are determined by the 
orientations of the core spins.  In the double-exchange picture 
[discussed in more detail in Sec.~(\ref{SEC:qp})], the rate $W_{ij}$ 
of carrier-hopping between ions $i$ and $j$, whose core spins (which we 
treat classically) form an angle $\theta$, is proportional to 
$\cos^2(\theta/2)$.  

\begin{figure}[hbt]
\epsfxsize=3in
\centerline{\epsfbox{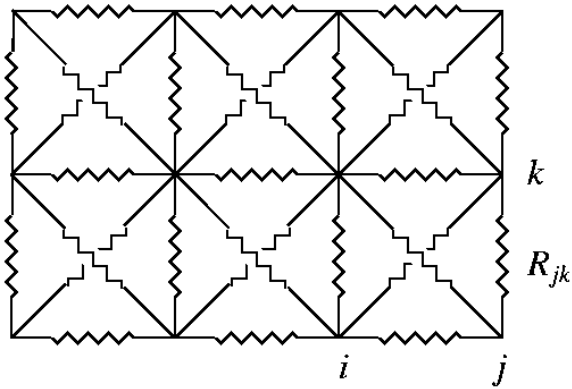}}
\vskip+0.20truecm
\caption{Schematic picture of the conducting network.
Zig-zag line connecting sites $i$ and $j$ represents 
the resistivity of the bond between these sites.}
\label{FIG:bondres}
\end{figure}%

As in the standard percolative approach to transport in impurity systems 
(i.e.~the Miller-Abrahams resistive-network 
approach~\cite{REF:AHL,REF:Efros,REF:Abr}), one connects every nearby pair 
of Mn ions by a bond, and assigns a resistivity to each of these bonds. 
The charge current $J_{ij}$ between ions $i$ and $j$ is then given by:
\begin{equation}
J_{ij} = e\left(W_{ij}-W_{ij}\right). 
\label{EQ:current}
\end{equation} 
In the presence of an applied electric field ${\bf E}$ and for a closed 
external circuit the system out of equilibrium and the charge current 
is nonzero.  If the electric field is sufficiently small 
(i.e.~$e{\bf E}\cdot({\bf R}_j -{\bf R}_i) \ll kT$), one can expand 
the charge carrier energies and site occupation numbers to linear 
order in the electric field.  Hence, one can obtain 
(see e.g.,~\cite{REF:Efros}) an expression for the electric current 
in Ohmic form: 
\begin{mathletters}
\begin{eqnarray}
J_{ij} &=& R^{-1}_{ij}(U_i - U_j),\\
R_{ij}&=&\frac{kT}{e^2 W^0_{ij}},
\label{EQ:Ohm}
\end{eqnarray}     
\end{mathletters}
where $U_i - U_j$ is the potential difference between sites $i$ and 
$j$ in the presence of the electric field, and $ W^0_{ij}$ is the 
$({\bf E}={\bf 0})$ transition rate.  Therefore, the resistance 
$R_{ij}$ of the bond between ions $i$ and $j$ (cf.~Fig.~\ref{FIG:bondres}) 
is proportional to $1/W_{ij}$, where we have, for the sake of brevity, 
omitted the superscript $0$ (which indicates zero electric fields 
quantities).  The resistances of the bonds constitute a resistive 
network on which carriers move by taking the least resistive paths, 
i.e., the transport is percolative in character.  The conductivity
of the sample is entirely determined by a set of hopping resistivities 
$R_{ij}$. 

\subsubsection{Scenario of transport properties in manganites}
\label{SEC:hopCP}

In the hopping regime, the following scenario for transport 
properties of manganites can be envisioned: 
\begin{figure}
\vskip0.6cm
\epsfig{figure=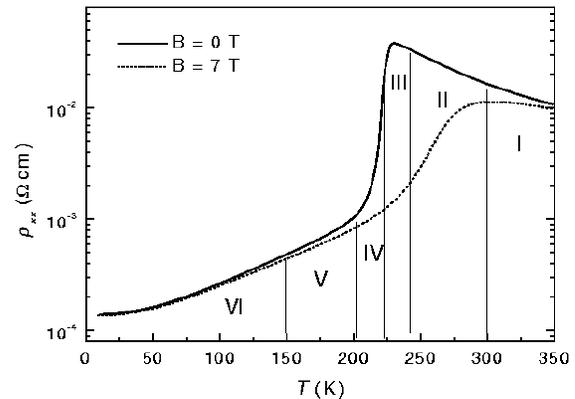,width=3.1in,rheight=3.0in,
angle=-90,silent=}    
\vskip-2.0cm
\caption{Transport regimes in manganites exeplified by LCMO.  
I:   High-temperature regime, in which transport is via polarons.  
II:  Crossover to inelastic hopping of charge-carriers between 
     localized states. 
III: Maximal resitivity.  In this regime the loss of 
     inelastic agents is compensated by the growth of magnetic 
     order and, hence, the emergence of a conducting network 
     (see text for details). 
IV:  Rapid growth of the conducting network. 
V:   Saturation of the conducting network. 
VI:  Crossover to the metallic regime.}
\label{FIG:tran}
\end{figure}   
\hfil\break\noindent
(i)~In the paramagnetic insulating state, i.e., in region~I on 
Fig.~\ref{FIG:tran}, the percolative inelastic motion 
of strongly localized carriers is suppressed by magnetic randomness, via 
the DE interaction.  Although there may exist small clusters of magnetic 
ions with spins aligned in some direction, neighboring spins (or the spins 
of neighboring aligned small clusters) are weakly correlated, and are 
therefore predominantly have a large angle between them.  Thus, the 
resistance of the corresponding resistive bonds is generally large. 
Hence, the clusters are isolated from each other (i.e.~no percolative 
path exists for which core spins on neighboring Mn ions are approximately 
aligned), so that outer-shell carriers cannot hop along any path of bonds 
without encountering a large resistance.  Furthermore, if the localization 
length is on the order of the lattice constant then carrier interactions 
with the lattice and carrier self-trapping via small-polaron formation 
become important.  (Below, we shall discuss the situation in which 
the localization length is larger than the lattice constant.)\thinspace\ 
Carrier self-trapping, when it occurs, does so on very weak
(i.e.~very resistive) bonds in the resistive network.  Deep in the insulating 
regime, all paths of bonds encounter regions in which carriers are 
self-trapped.  Under these conditions, transport occurs via the rather 
infrequent hopping of small polarons, which leads to the thermally-activated 
temperature-dependence of the longitudinal resistivity and Hall coefficient. 
In the LCMO and LPCMO compounds, the role of polarons in transport accounts 
for the magnitude of the activation energy associated with the resistivity, 
which is above $100\,{\rm meV}$, and is significantly larger than (any 
fraction of) the hopping amplitude $t$.   
\hfil\break\noindent
(ii)~With decreasing temperature, the inelastic hopping becomes less 
frequent, so that the resistivity grows.  However, at these temperatures, 
percolative paths appear that do not encounter regions with self-trapped 
carriers.  Also, the core spins become more aligned with one another, 
and fairly large clusters exist (that do not feature large resistances  
connecting pairs of Mn ions having anti-aligned spins). This regime 
occurs in the range~II of temperatures in Fig.~\ref{FIG:tran}.  
\hfil\break\noindent
(iii)~With continued decrease in temperature, the resistivity reaches 
a maximum (region~III in Fig.~\ref{FIG:tran}) 
when the core spin orientations become sufficiently correlated 
that a tenuous but infinite conducting network emerges.  Due to the 
still-strong magnetic disorder, as well as any nonmagnetic disorder, the 
carrier states are still localized (and lie in the band tail). The 
localization length is on the order of one to two Mn sublattice units. 
(It is important to note that clusters of spins of size two sublattice 
units contain some 20 to 30 spins.  The alignment of these spins can be 
obtained by applying magnetic fields of order only a few Tesla, leading 
to colossal magnetoresistance.) 
\hfil\break\noindent
(iv)~With yet further reduction in temperature, the resistivity decreases 
abruptly, in line with the standard percolation 
picture~\cite{REF:Efros,REF:AHL}, 
as more and more inelastic hopping-paths become available to carriers, 
owing to the increased alignment of core spins (region~IV in 
Fig.~\ref{FIG:tran}).  The small polarons 
disappear and, in addition, some of what used to be localized states 
become delocalized, so that some carriers now populate the states lying 
on the mobile (i.e.~metallic conduction) side of the mobility threshold.  
(v)~The abrupt decrease in resistivity with temperature slows down as soon as 
any newly-available hopping paths are effectively shunted by the existing 
conducting network (region~V in Fig.~\ref{FIG:tran}). 
\hfil\break\noindent
(vi)~Further decrease in temperature leads to further core-spin alignment 
and, ultimately, to a significant density of carriers populating states 
in the conducting part of the band and, hence, to the occurrence of 
the metallic state (region~VI in Fig.~\ref{FIG:tran}). 

\subsubsection{Scenario of transport properties of manganites: 
Absence and Presence of polarons.}
\label{SEC:hopS}

As described in Sec.~\ref{SEC:expres}, the temperature dependence of 
the resistivity does not have a universal form across all manganite 
compounds.  In particular, in LCMO and LPMO, the high-temperature 
$\rho_{xx}$ is thermally activated, but in LSMO it is not.  It is our 
opinion that the scenario described in points~(i-vi) in the previous 
paragraph, which involves self-trapping effects due to polaron 
formation, takes place in LCMO and LPMO but not in LSMO. 
This opinion is supported not only by our transport data  
but also by direct neutron-scattering evidence for the coexistence of 
distorted and undistorted Mn-O octahedra in the vicinity of 
$T_{C}$~\cite{Louca,Billinge,Lynn}, as well as by the occurrence of a 
substantial isotope effect in the resistivity, $T_C$ and thermal 
expansion~\cite{isotope}.  

By contrast, LSMO shows no evidence of polarons in this suite of 
experiments.  Therefore we propose that in LSMO it is only magnetic 
and static disorder that drive the transition between low- and 
high-resistance states.  According to the scenario described in the 
previous paragraph, any tendency for the formation of polarons 
is suppressed when the localization length at the resistive transition 
(i.e.~the inflection point in LSMO) turns out to be larger than a few 
lattice constants.  Because of this, with further increase in the 
temperature the localization length still has room to decrease due to  
the suppression of ferromagnetic correlations.  This would lead to an 
increase of the resistivity at temperatures in the immediate vicinity 
above the resistive transition. Such a resistivity increase has been 
observed in LSMO. 

The significance of polarons in LCMO and LPMO, and their apparent 
insignificance in LSMO, is consistent with the tendency for polaronic 
self-trapping to be enhanced for reduced A-site ionic radius, as is
the case in the sequence Sr$\rightarrow$Pb$\rightarrow$Ca encountered 
in our experimental data.  The significance of polarons in LCMO and 
LPMO can also explain why the magnetoresistance of these compounds is 
stronger than that of LSMO: In LCMO and LPMO, the application of a 
magnetic field not only results in the tendency to delocalize charge 
carriers by reducing the magnetic disorder, as it does in LSMO, but 
also such magnetic fields destabilize the polaronic regions, leading 
to a more abrupt reconnection of the network. Furthermore, the 
presence of polaronic and non-polaronic spatial regions in LCMO and 
LPMO enhances disconnection between different parts of the resistive 
network, and, due to the DE origin of ferromagnetism in these 
systems, reduces the transition temperature. This 
\lq\lq boot-strap\rq\rq\ collapsing of magnetic order explains 
why the bare double-exchange energy that determined the spin-wave 
dispersion at low temperatures is the same for all 
manganites~\cite{spinwave}, even though 
$T_{\rm C}^{\rm LCMO}=220$K, 
$T_{\rm C}^{\rm LPMO}=285$K, and 
$T_{\rm C}^{\rm LSMO}=360$K.  
We propose that the different sizes of the A-site ions which, in 
particular, leads to differing-strengths in the static disorder 
and differing tendencies for self-trapping, is responsible for this 
trend in the transition temperatures.  We also note, that the 
existence of polaronic and non-polaronic spatial regions in LCMO 
and LPMO explains the success of the effective medium approaches 
in predicting the thermoelectric power from the 
resistivity~\cite{REF:Jaime,thermopower}, the observation of both diffusive 
and continuum electronic signals in Raman scattering~\cite{Raman}, 
and the presence of significant telegraph noise in the resistivity 
in these compounds~\cite{mweissman}. These features are not 
characteristic for LSMO.

In Secs.~\ref{SEC:hophall}-\ref{SEC:structure} and \ref{SEC:disctheorex}, 
we shall look at the scenario that 
we have just outlined from the vantage point afforded by the Hall 
effect.  As, in our opinion, the dominant transport mechanism in the 
transition regime occurs via inelastic hopping between localized 
states, regardless of whether self-trapping via the formation of small 
polarons occurs, in sections~
\ref{SEC:holst} we shall first review the basic physical picture 
underlying the ordinary Hall effect in the inelastic hopping regime 
in systems having potential (but not magnetic) disorder.  The main issues 
of this discussion are interference of hopping amplitudes in the 
inelastic hopping regime and elucidation of contribution to the 
Hall effect in this regime by using properties of the hopping probability 
with respect to time-reversal symmetry 
Secs.~\ref{SEC:inel} - \ref{SEC:OHEinterferenceP}. 
(We obtain expressions 
for hopping amplitudes which differ from those of Holstein 
in~\cite{REF:Holstein1}, but this difference is not significant). 
Readers familiar with these  
issues and Holstein theory of the Hall effect can proceed to subsections 
following this discussion, in which 
we shall provide an extended discussion of our picture of the 
microscopic mechanism of the anomalous Hall effect in the hopping 
regime in manganites, which we have recently proposed~\cite{LPMO-PRL,Yuli}.    

\subsection{Holstein theory of the Hall effect in the hopping regime}
\label{SEC:holst}

Nearly forty years ago, Holstein~\cite{REF:Holstein1} observed 
that to capture the ordinary Hall effect in hopping conductors 
requires the analysis of {\it at least triads\/} of sites 
(i.e.~atoms, ions, impurities, etc.), and of the attendant 
Aharonov-Bohm (AB) magnetic fluxes through the polygons whose vertices are 
these sites.  What Holstein showed was that the probability of 
the hopping of a charge carrier that is initially located on one 
of three sites $i$ to one or the other of the remaining sites $j$ (which 
are initially assumed to be unoccupied), $W_{ij}$ contains a contribution 
$\delta W^{\rm H}_{ij}$ that
\begin{figure}[hbt]
 \epsfxsize=2.2in
\centerline{\epsfbox{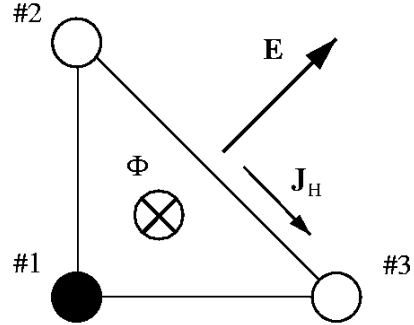}}
\vskip0.4cm
\caption{A triad of sites is the minimal element of the 
conducting network that leads to a Hall electromotive force. 
Owing to the flux $\Phi$ the carrier current ${\bf J}_{\rm H}$ 
flows perpendicular to the electric field ${\bf E}$.  
The shaded site is occupied; the unshaded sites are unoccupied.}
\label{FIG:htr}
\end{figure}%
\noindent
is linear in the applied magnetic field.  This dependence  
arises from interference between the amplitude for a direct 
(inelastic) transition between the initial and final sites and 
the amplitude for an indirect (inelastic) transition, involving 
the intermediate occupation of the third site $k$. 
Furthermore, when one applies a magnetic field creating a 
nonzero magnetic flux through the triangle, thus 
introducing an Aharonov-Bohm phase for paths that wind around 
the triangle, the necessity that  $W_{ij}$ and $W_{ji}$ be equal is lost, 
even if ${\bf E}\perp ({\bf R}_i -{\bf R}_j) $.
The Hall current, flowing through the bond between sites 
$i$ and $j$ in a triad of sites, 
in the presence of such electric field, and in magnetic
field perpendicular to the plane containing sites  $i$, $j$ and $k$,  
as sketched in Fig.~\ref{FIG:htr}, 
is given by [c.f.~Eq.~(\ref{EQ:current})]
\begin{equation}
J^{\rm H}_{ij} = e\left(\delta W^{\rm H}_{ij}-\delta W^{\rm H}_{ji}\right).
\label{EQ:hallcur}
\end{equation}
Such current would cause an increasing imbalance of 
populations of sites $i$ and $j$. However, the balance is restored 
because charge imbalance establishes a chemical potential 
difference $\Delta\mu _{ij}$ between sites~$j$ and $i$ which manifests 
itself as a Hall voltage. 
Below in Secs.~\ref{SEC:hophall} - \ref{SEC:structure} 
we shall generalize this idea to 
charge carrier motion in the presence of core spins having 
inhomogeneous orientations.

\subsubsection{Ordinary Hall effect: Direct and indirect hopping}
\label{SEC:caricature}

Let us now address the issue of the Hall effect at an elementary 
level.  To do this we consider a system of localized carriers, 
their wavefunctions $\ipr{{\bf r}}{\Psi_j}$ ($j=1,2,\ldots$) being 
the exact wave functions of the discrete spectrum of the electronic 
Hamiltonian $H_{\rm el}$ in the presence of ionic potentials, 
potential disorder and magnetic field.  (In order to be specific, 
we assume, in the present subsection, that the carriers are 
electrons.)

Now consider the rates of hopping between these exact electronic 
states caused by the electron-phonon interaction $W_{j\rightarrow k}$. 
The necessity of the electron-phonon interaction (or interactions 
with some other inelastic agent) for inducing electron hopping will 
be discussed below in the present subsection.  Within the context of 
hopping rates it is valuable to introduce the notion of {\it direct\/} 
and {\it indirect\/} hopping rates.  The direct hopping rate 
$W^{\rm dir}_{j\rightarrow k}$ is, to leading order in the electron-phonon 
interaction, determined by the single direct transition amplitude, i.e., 
the electron-phonon interaction matrix element 
\begin{equation}
U_{jk}=\langle\Psi_j\vert H_{\rm el-ph}\vert\Psi_k\rangle,
\label{EQ:wantME}
\end{equation}
where $H_{\rm el-ph}$ is the Hamiltonian of electron-phonon interaction.  
In Fermi's Golden Rule approximation $W^{\rm dir}_{j\rightarrow k}$ reads
\begin{equation}  
W^{\rm dir}_{j\rightarrow k}=
\frac{2\pi}{\hbar}\vert U_{jk}\vert^2 
\delta(E_j-E_k-\hbar\omega),
\label{EQ:FGRrate}
\end{equation}
where $\{E_i\}$ are the exact energy eigenvalues of  $H_{\rm el}$, 
and $\omega$ is a phonon frequency.  There are additive corrections 
to the direct transition amplitude, which are associated with processes 
involving phonon-induced scatterings that do not change the 
electronic state.  In addition to the direct transition amplitude, 
there are amplitudes for indirect transitions from site $j$ to site 
$k$, which are defined to be those amplitudes that involve at least 
one intermediate eigenstate $\ket{\Psi_i}$ (but now $i$ is 
restricted to be neither $j$ nor $k$).  Among these, there is a 
subset of amplitudes involving exactly one intermediate eigenstate.  
Such amplitudes have the following characteristic property: The 
indirect amplitude that proceeds via a third site $i$ necessarily 
involves two electron-phonon interaction matrix elements 
$U_{ij}$ and $U_{ki}$.  Direct and indirect transition amplitudes can 
interfere, and, as we shall describe below, lead to the Hall effect.

\subsubsection{Hopping transport: compatibility of 
inelastic processes and quantum interference}
\label{SEC:inel}

Despite the apparent simplicity of the foregoing analysis of a 
triad of sites, the task of obtaining the linear dependence 
of the Hall resistivity on the magnetic field via Holstein's 
approach is a much more subtle matter.  Furthermore, the 
issue of establishing quantal interference effects in this 
setting of inelastic hopping processes is equally subtle, 
so we shall now revisit this subject.
\begin{figure}[hbt]
  \epsfxsize=2.5in
\centerline{\epsfbox{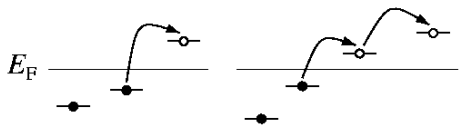}}
\vskip+0.20truecm
\caption{Schematic depiction of phonon-assisted inelastic 
hopping.  Direct hopping (left) and indirect coherent 
hopping via an intermediate site (right) are shown.}
\label{FIG:inelahop}
\end{figure}%

Why is it that we need to consider inelastic processes when 
considering the Hall effect?  As for the longitudinal hopping 
conductivity of localized carriers, it is due to electronic 
quantum transitions between localized carrier eigenstates, 
assisted by phonons (or some other inelastic agent).  The 
participation of some inelastic agent in hopping conduction is 
required for the following reasons.  First, owing to carrier 
localization, the conductivity of the carrier system in the 
absence of phonons is that of an Anderson insulator, i.e., zero.  
Phonons cause transitions between localized carrier states and, 
hence, allow conduction.  Second, phonon-assisted carrier 
transitions meet the need to have carrier transitions between 
occupied states (which lie below the Fermi level) and unoccupied 
states (which lie above the Fermi level) whilst satisfying the 
demand that energy be conserved (Fig.~\ref{FIG:inelahop}).  This energy-conservation 
requirement also holds for the Hall effect.  Now, as inelastic 
processes are being invoked  one should ask the question:  How 
can interference between distinct hopping paths, necessary for 
the sensitivity of the hopping rate to any quantal phase, arise?  

We will answer this question both in the context of longitudinal 
conductivity and the Hall effect. In the next subsubsection we 
will consider the longitudinal conductivity. After that, in further 
sections,  
we will discuss what inelastic processes contribute to the Hall effect, 
and interference of those processes.

\subsubsection{Quantum interference in the presence of inelastic scattering.
Longitudinal conductivity.}
\label{SEC:OHEinterferenceL}
 
In order to answer the question posed in the previous section 
let us first formulate precisely what 
is meant by coherence and sensitivity to the quantal phase in the 
system at hand. The Hamiltonian of the system reads
\begin{equation}
H= H_{\rm el} + H_{\rm el-ph} + H_{\rm ph},
\end{equation}
where $ H_{\rm ph}$ is the phonon Hamiltonian.  The exact
quantum-mechanical eigenstates of  $H_{\rm el}$, both localized 
and delocalized, are sensitive to applied magnetic field. 
(Here, we are, of course, concerned with localized states.)\thinspace\ 
In the presence of a magnetic field, the exact eigenfunctions  
$\ipr{{\bf r}}{\Psi_j}$ can no longer be chosen to be real quantities, 
and are thus characterized by an absolute value and a phase.  
It is often convenient to approximate the exact localized eigenstates 
in terms of the eigenstates $\{\ket{\phi_j}\}$ of outer-shell electrons 
of isolated ions (where $i$ enumerates the ions).  We note that the 
$\{\ket{\phi_j}\}$ are not, in general, orthogonal, although they 
are typically linearly independent and may, therefore, be taken as 
a basis.  In terms of this basis a Hamiltonian describing carrier 
motion in the presence of the corresponding isolated ions located at 
positions $\{{\bf R}_j\}$ as well as a disordered potential has the 
form 
\begin{equation}
H_{\rm el} =
\sum\limits_j
\ket{\phi_j}\epsilon_j\bra{\phi_j} 
+\sum\limits_{j\ne k}
\ket{\phi_j}V_{jk}\bra{\phi_k}, 
\label{EQ:el}
\end{equation}
where $\{\epsilon_j\}$ are random energies and $V_{jk}$ are 
random transfer matrix elements.  

Having introduced the basis of localized ionic states 
$\{\ket{\phi_{j}}\}$ and the Hamiltonian for the disordered 
system of ions $H_{\rm el}$, we now examine in detail 
the effect on this system of an applied magnetic field. 
As shown by Holstein~\cite{REF:Holstein1}, in the presence of a 
magnetic field the ionic basis states $\{\vert\phi_j\rangle\}$ 
become the modified collection 
$\{\vert\phi_j\rangle_{B}\}$, 
being solutions of the Schr\"odinger equation
\begin{equation}
\left[\frac{1}{2m^*}
\left({\bf p}+\frac{e{\bf A}}{c}\right)^2 - 
U({\bf r}- {\bf R}_j)\right]
\ipr{{\bf r}}{\phi_j}_{B}= 
\epsilon
\ipr{{\bf r}}{\phi_j}_{B},  
\label{mods}
\end{equation}
where $U({\bf r}-{\bf R}_j)$ is the potential of the ion 
located at position ${\bf R}_j$, $m^*$ is the effective mass, 
and ${\bf A}$ is the magnetic vector potential.  The 
wave function $\ipr{{\bf r}}{\phi_j}_{B}\vert_{B=0}$ has the 
property that it is a function of ${\bf r}-{\bf R}_j$, and we 
would like to recover something like this property in the 
presence of the magnetic field.  To do this we introduce a 
gauge transformation
\begin{mathletters} 
\begin{eqnarray}
\ipr{{\bf r}}{\phi_j}
&=& 
\ipr{{\bf r}}{\kappa_j}\, 
e^{-i\eta_j({\bf r})},
\label{gauge}
\\
\eta_j({\bf r})
&=& 
(e/\hbar c)
{\bf A}({\bf R}_j)\cdot{\bf r}= 
(e/\hbar c)
({\bf B}\times{\bf R}_j)\cdot{\bf r},
\end{eqnarray}
\end{mathletters}
where we have chosen the gauge potential to be 
${\bf A}({\bf r})={\bf H}\times{\bf r}/2$.
It is straightforward to show that this transformation 
leads to an equation for $\ipr{{\bf r}}{\kappa_j}$ in  
which the vector potential (magnetic field) term contains 
coordinate in combination $({\bf r}-{\bf R}_j)$:
\begin{eqnarray}
\left[\frac{1}{2m^*}
\left({\bf p}+e{\bf H}\times({\bf r}-{\bf R}_j)/2c\right)^2 
\right.\nonumber\\
\qquad\qquad
\left.- U({\bf r}- {\bf R}_j)\right]
\ipr{{\bf r}}{\phi_j}_{B}= 
\epsilon
\ipr{{\bf r}}{\phi_j}_{B},  
\label{locS}
\end{eqnarray}
Hence, the 
wave function $\ipr{{\bf r}}{\kappa_j}$ is seen to have 
the sought property: 
\begin{equation}
\ipr{{\bf r}}{\kappa_j}_{B}=
\ipr{{\bf r}-{\bf R}_j}{\kappa_j}.
\end{equation}
It is convenient to employ the magnetic-field dependent ionic 
states $\{\ket{\phi_i}_{B}\}$ in the perturbative construction of 
the eigenstates of the system Hamiltonian~(\ref{EQ:el})
\begin{equation}
H_{\rm el}({\bf B})\!=\!
\sum\limits_j
\ket{\phi_j}_{B}\,
\epsilon_{j}({\bf B})\,
\bra{\phi_j}_{B} 
+\sum\limits_{j\ne k}
\ket{\phi_j}_{B}\,
V_{jk}({\bf B})\,
\bra{\phi_k}_{B}.
\label{EQ:ElWithB}
\end{equation}
In what follows we shall omit the explicit dependence on ${\bf B}$.

We now construct approximations to the exact eigenstates of 
$H_{\rm el}({\bf B})$ in terms of linear combinations of ionic 
states $\{\ket{\phi_i}_{B}\}$.  To do this we use renormalized 
(i.e.~Brillouin-Wigner) perturbation theory in powers of 
$V_{kj}/(E_j - E_k)$ (see, e.g., 
Ref.~\cite{REF:Ziman,REF:Schirmacher}), thus obtaining 
\begin{eqnarray}
\ket{\Psi_j}
&=&
\ket{\phi_j}+
\sum\limits_{k(\ne j)}
\ket{\phi_k} 
\left[
\frac{V_{kj}}{E_j - E_k}
\right.
\nonumber\\ 
&& 
\qquad
\left. 
+\sum\limits_{k(\ne j)\atop{h(\ne j)}}
\frac{V_{kh}V_{hj}}{(E_j - E_k)(E_j - E_h)}
+\cdots
\right],
\label{expansion}
\end{eqnarray}
where $\{E_j\}$ are the exact energy eigenvalues. 
We now pause to remark that in the theory of hopping conductivity 
in doped semiconductors~\cite{REF:Abr,REF:Efros,REF:AHL,REF:Holstein1} 
the parameter $V_{kj}/(E_j - E_k)$ 
is indeed small, due to sizable 
distance between donors (or acceptors). In subsection 
devoted to application of hopping conductivity model to manganites we 
we will see that 
this parameter can also be rendered as small, because 
of the dependence of the effective hopping amplitude on the core-spin 
misalignment. It is worth mentioning, however, that
 the  
sensitivity of $\ket{\Psi_j}$ to phases arising from transformation 
Eq.~(\ref{gauge}) is a general property which 
does not rely on perturbation expansion~\ref{expansion}.
It is also reasonable to assert that the 
essential dependence of $\{\ket{\Psi}_j\}$ on the magnetic 
field enters solely through such phase acquired by the local 
basis wave functions $\{\ipr{{\bf r}}{\phi_j}\}$ under the 
gauge transformation in Eq.~(\ref{gauge}), because, 
at reasonable experimental strengths of the magnetic
field much less than a quantum of flux treads a localized
ionic wave function.

\eqbreak

Having constructed the states $\ket{\Psi_j}$, i.e., approximations 
to the exact localized states, we now use them to compute the square 
modulus of the matrix element $U_{jk}$ of Eq.~(\ref{EQ:wantME}): 
\begin{eqnarray}
\vert{U_{jk}}\vert^{2}
&=&
\vert{\langle\Psi_j\vert H_{\rm el-ph}\vert\Psi_k\rangle}\vert^{2}
\nonumber\\
&=&
\cdots+
\sum\limits_{i\ne k}
\sum\limits_{n\ne k, n\ne i}
\me{\phi_k}{H_{\rm el-ph}}{\phi_j}
\me{\phi_j}{H_{\rm el-ph}}{\phi_k}
\frac{V_{ik}V_{kn}V_{ni}}{(E_i-E_k)^2(E_i-E_n)}
\nonumber\\
&&
\phantom{\cdots}
+\sum\limits_{m\ne j}
\sum\limits_{l\ne m, l\ne j}
\me{\phi_i}{H_{\rm el-ph}}{\phi_m}
\me{\phi_m}{H_{\rm el-ph}}{\phi_i}
\frac{V_{mj}V_{jl}V_{lm}}{(E_j-E_m)^2(E_j-E_l)}+
{\rm c.c.}+\cdots, 
\label{EQ:MEeval}
\end{eqnarray}
which features in the transition rate, Eq.~(\ref{EQ:FGRrate}). 

\eqresume

Terms exhibited in Eq.~(\ref{EQ:MEeval}) involve motion along 
paths that surround loops of nonzero area, i.e., the matrix elements 
of electron-phonon interaction and transfer amplitudes 
are taken between localized orbitals Eq.~(\ref{mods}) of sites that form 
such loops. Note, however, that not all such terms are included 
in Eq.~(\ref{EQ:MEeval}), but only those in which matrix elements 
of electron-phonon interaction enter the corresponding expressions 
in combination with their complex conjugated (i.e. time-reversed) 
counterparts. Furthermore, the remaining matrix elements entering 
terms featured in Eq.~(\ref{EQ:MEeval}), i.e., the overlap integrals $V_{hk}$, 
 involve motion along paths that surround loops of nonzero area~\cite{RB}.   
It is through such products of three overlap integrals $V_{hk}V_{kj}V_{jh}$ 
that the transition rate acquires its sensitivity to fluxes 
through loops of nonzero area.
It is only such terms that lead to non-vanishing interference 
contribution to hopping probability which is sensitive to fluxes. 
Such sensitivity to phase results in the Aharonov-Bohm magnetoresistance 
effects in hopping conductivity (see, e.g., Ref.~\cite{Spivak}).

Let us discuss the physical meaning of interference terms featured in 
Eq.~(\ref{EQ:MEeval}).
Consider, for example, charge carrier hopping in a triad of sites. 
The relevant terms  in Eq.~(\ref{EQ:MEeval}), which involve loops, 
correspond to $i=1$, $j=2$, $k=2$, $n=3$ (in the first of featured terms), 
and $m=1$, $l=3$ 
(in the second of featured terms). In terms of isolated orbitals 
the first term features in~(\ref{EQ:MEeval}) corresponds to 
interference of two processes of 
charge carrier tunneling from site 1 to site 2, a direct one 
and an indirect one (via site 3), 
with the following charge carrier interaction with a phonon at site 2; 
the second term corresponds to carrier interaction with phonon at site 1 
with the following interference between direct and indirect tunneling paths 
from site 1 to site 2.
(We note that the situation can, 
of course, be readily generalized to the case  of changes in more 
than one phonon occupation number.) 

We now notice that by contrast with terms that are featured 
in~(\ref{EQ:MEeval}), terms that 
have been omitted there, like 
\begin{equation}
\delta\vert{U_{jk}}\vert^{2}=\sum\limits_{k\ne i}
\me{\phi_i}{H_{\rm el-ph}}{\phi_j}\me{\phi_j}{H_{\rm el-ph}}{\phi_k}
\frac{V_{ki}}{E_i-E_k},
\end{equation}
involve motion along paths that surround loops of nonzero area, but 
do not lead to non-vanishing interference contribution to hopping probability. 
The reason for absence of such contributions is that they 
do not involve combinations of matrix elements of electron-phonon 
interaction with their time-reversed counterparts. 
To see why such combinations are important, consider the 
electron-phonon interaction Hamiltonian
 \begin{equation}
H_{\rm el-ph}=\sum_{\bf q} H_{q}
e^{i{\bf q}\cdot{\bf r}},
\label{EQ:elph}
\end{equation}
where ${\bf q}$ is the phonon wavevector. Evaluation of non-vanishing 
interference contribution, therefore, includes sums over all 
phonon wavevectors. Terms that exhibit two matrix elements of 
electron-phonon interaction that are not complex conjugated to each other 
turn out to be oscillating functions of ${\bf q}$, and vanish upon 
evaluation of the sum over  ${\bf q}$. This simple observation allows us 
to find all important interference terms in the hopping probability 
~(\ref{EQ:MEeval} without resorting to explicit evaluation of 
$\me{\phi_i}{H_{\rm el-ph}}{\phi_j}$ as it was done, e.g., in 
~\cite{REF:Holstein1} in consideration of the Hall effect and in
\cite{REF:Schirmacher} in consideration of the Aharonov-Bohn hopping 
magnetoresistance. We note that if the sum of two terms featured in 
~(\ref{EQ:MEeval}) is equal to zero, this consideration easily allows us to
find the appropriate next terms of expansion  of hopping probability 
in powers of $V_{ij}/(E_i-E_j)$.

We are now in a position to discuss what is meant by phase coherence 
and sensitivity to quantal phases in the present setting of hopping 
conduction.  In the absence of inelastic agents, 
products of, e.g., three overlap integrals $V_{hk}V_{kj}V_{jh}$ (where sites 
$h$, $k$ and $j$ form a nondegenerate triangle, with non-zero flux threading
this triangle in the presence of magnetic field), and 
carrier energies $E_j$ (see the remark \cite{RB}) are sensitive 
to the Aharonov-Bohm quantal phase. As we can see from Eq.~(\ref{EQ:MEeval}), 
interference of two amplitudes of quantal transition between states 
$h$ and $k$ occurs (and these amplitudes are coherent) 
even if these transitions are due to 
inelastic agents. For such interference to occur, inelastic agents which 
determine one of the amplitudes of the transition must be the same as 
inelastic agents which determine another amplitude of the transition 
(i.e. phonon frequencies and wavevectors are equal), so that the 
square modulus of   
matrix element of, e.g., the electron-phonon interaction, enters the 
probability. As we have mentioned above, the presence of square modulus of the 
electron-phonon interaction matrix element in the probability 
corresponds to phonon factors in the two interfering amplitudes 
(e.g., in the amplitude of a direct process and in the 
complex conjugate of the amplitude of an indirect process) being related 
to each other by the time-reversal. 

Let us now discuss what is meant by the phase breaking or decoherence in the 
context of the hopping conductivity regime. The eigenstates $\{\ket{\Psi}_j\}$ 
and their energies $E_j$ are determined by $H_{\rm el-ph}$, i.e., 
in the absence of phonons. When phonons interact with charge carriers, 
one can consider them as a reservoir 
which leads to randomization of carrier states. The time scale of such
randomization is given by the inverse rate of carrier transitions  between 
different eigenstates of $H_{\rm el-ph}$ caused by electron-phonon 
interaction (\ref{EQ:FGRrate}). It is this randomization that is meant by 
phase breaking (or decoherence). However, the carrier transition rate 
(\ref{EQ:FGRrate}) between different states itself, which measures the rate of 
decoherence (being determined by 
eigenfunctions of these states) carries information about quantal phases 
that arise in the context of Eqs.~(\ref{expansion}) and ~(\ref{EQ:MEeval}).
Furthermore, the electric current arises during the act of hopping, i.e. 
when randomization has not yet occurred.   
Therefore, the answer to the question, posed in the second paragraph 
of the present subsection, concerning interference in the presence 
of inelastically-assisted hopping is as follows: The (steady-state) 
hopping current is generated during the process of inelastic scattering, 
whilst decoherence arises only after this scattering has occurred;
thus, decoherence effects do not preclude sensitivity of hopping conduction 
to quantal phases.  
Moreover, the question of interference of consequent hopping
events is meaningless in the context of hopping conductivity, 
because only amplitudes of hopping between the same initial and the same 
final states can interfere. Thus, there is a significant difference 
between interference effects in hopping conductivity and in diffusive 
mesoscopic transport. In diffusive transport, the whole diffusive trajectory, 
with all consequent scattering events, determines the current via 
the diffusivity, and electronic coherence during consequent scattering 
events is important for interference effects. In hopping transport, 
incoherence of consequent hopping events does not contradict 
interference of quantum-mechanical amplitudes that determine 
a single charge carrier hop and the hopping current itself.

\subsubsection{Quantum interference in the presence of 
inelastic scattering; processes leading to the Hall effect}
\label{SEC:OHEinterferenceP}

For localized carriers, the interference processes that lead to 
the Hall conductivity are even more peculiar than those leading 
to the sensitivity of the longitudinal conductivity to the 
Aharonov-Bohm phase.  In particular, the occurrence of the Hall 
effect requires phonon-assisted hoppings between exact initial 
and final carrier states, via an exact intermediate state.  (It 
is not sufficient to include as intermediate states the virtual 
ionic orbitals that enter via Eq.~(\ref{expansion}.)\thinspace\  
The reason for this difference between the hopping Hall conductance 
and the sensitivity to magnetic flux of the hopping 
magnetoconductance arises because of the necessity to extract a 
dependence of the Hall conductance that is {\it linear\/} in (or, 
more generally, an odd power of) the magnetic field .

Let us now explore the collection of processes that contribute to 
the Hall conductance.  These processes, first identified by Holstein, 
can be determined by making use of the odd (i.e.~dissipative) 
character of transition rates under the time-reversal operation: 
$t\to -t$.  As discussed in Sec.~\ref{SEC:Perc_scen} and \ref{SEC:holst}, 
the hopping current 
is determined, via the conductivities of the resistive network, 
by the rate of transitions between sites.  The Hall current, as 
with any current, is odd under time-reversal.  So, too, is the 
transition rate. (To appreciate this, consider an elementary 
relaxation process for some quantity $Q$, which evolves according to 
the rate equation 
\begin{equation}
\frac{dQ}{dt} \equiv -\frac{Q}{\tau}.
\end{equation}
From the consistency of equation one immediately sees that it 
is formally appropriate to designate the relaxation rate $\tau$ 
as being odd under the time reversal.)\thinspace\ Now consider the 
transition probability per unit time between the exact single-carrier 
states $i$ and $f$, viz.~$W_{fi}$ which, according to Fermi's Golden 
Rule,  is given by 
\begin{equation}
W_{fi}=\frac{2\pi}{\hbar}\vert A_{fi}\vert^2\,\delta(E_i - E_f),
\label{EQ:transrate}
\end{equation}
where $A_{fi}$ is the sum of the transition amplitudes for all 
coherent (i.e.~interfering) processes connecting $i$ and $f$, and 
the $\delta$ function imposes energy conservation between the 
initial and final states, the energies $E_i$ and $E_f$ of these 
states being full energies. (That is, they include not only carrier 
but also phonon energies).  The $\delta$-function, which has the 
complex representation
\begin{equation}
\delta(E)=
{\rm Im}\frac{1}{\pi}\lim_{s\rightarrow +0} 
\frac{i}{E-is}, 
\label{prob}
\end{equation}
is an odd quantity with regard to time-reversal, in the sense that 
under the transformation $t\rightarrow -t$, sign of imaginary part 
$s$ in the denominator changes and, thus, so does the sign of the 
$\delta$ function.  By contrast, the quantity $\vert A_{fi}\vert^2$ 
is even (i.e.~non-dissipative) under time-reversal.  We note, in 
passing, that the precision of the energy conservation is not what 
is essential (from the point of view of time-reversal properties). 
For example, the imaginary part of a Lorentzian function, 
\begin{equation}
{\rm Im}\,
\frac{1}{\pi}
\frac{i}{E-i\Gamma}=
\frac{1}{\pi}
\frac{\Gamma}{E^{2}+\Gamma^{2}}, 
\label{EQ:Lorentz}
\end{equation}
reveals that this function, too, inherits the oddness of the rate $\Gamma$.

En route to exploring the processes that contribute to the Hall 
conductance, let us now consider a simple case in which $A_{fi}$ 
has just two contributions: 
\begin{equation}
A_{fi}=A^{\rm dir}_{fi} + A^{\rm ind}_{fi}, 
\end{equation}
where $A^{\rm dir}_{fi}$ is the amplitude for the direct path and 
$A^{\rm ind}_{fi}$ is the amplitude for an indirect path (i.e.~a 
path via an intermediate state).  These two amplitudes can be 
written in the form
\begin{mathletters}
\begin{eqnarray}
A^{\rm dir}_{fi}& = & A^{0,{\rm dir}}_{fi}\exp{i\phi_1},\\
A^{\rm ind}_{fi}& = & A^{0,{\rm ind}}_{fi}\exp{i\phi_2},\\
\phi&=&\phi_1-\phi_2, 
\end{eqnarray}%
\end{mathletters}%
where 
$A^{0,{\rm dir}}_{fi}$ and 
$A^{0,{\rm dir}}_{fi}$ are the zero-magnetic-field amplitudes, and 
$\phi_1$ and $\phi_2$ are phases arising in the presence of magnetic 
field for direct and indirect paths, correspondingly, and $\phi$ is the 
difference of phases between direct and indirect paths induced by 
magnetic field, i.e., the Aharonov-Bohm phase.  (We neglect changes 
magnetic-field-induced changes in the magnitudes of 
$A^{0,{\rm dir}}_{fi}$ and 
$A^{0,{\rm dir}}_{fi}$.)\thinspace\ 
For the transition probability we then have
\begin{eqnarray}
\vert A_{fi}\vert^2 &=& 
 \vert A^{0,{\rm dir}}_{fi}\vert^2  
+\vert A^{0,{\rm ind}}_{fi}\vert^2 +
{\rm Re}\,
A^{0,{\rm dir}\,\ast}_{fi} 
A^{0,{\rm ind}}_{fi}\,\cos\phi
\nonumber\\ 
&&\qquad\qquad
+{\rm Im}\,
A^{0,{\rm dir}\,\ast}_{fi}
A^{0,{\rm ind}}_{fi}\,\sin\phi.
\label{EQ:sinphi}
\end{eqnarray}
We now observe that the term containing $\sin{\phi}$ is the only 
contribution to the probability $\vert A_{fi}\vert^2$ that is odd 
with respect to the transformation $\phi\rightarrow-\phi$ (i.e.~with 
respect to magnetic-field reversal), this oddness being a necessary 
property of the Hall conductance.  Thus, in a computation of the Hall 
conductance, only the imaginary part of the quantity corresponding to 
$A^{0,{\rm dir}\,\ast}_{fi}A^{0,{\rm ind}}_{fi}$ contributes and, 
therefore, one has to consider those indirect processes for which 
the zero-magnetic-field amplitude has a component out-of-phase with
the zero-magnetic-field amplitude of the direct process.  (Said another 
way, one must consider contributions 
to $A^{0,{\rm dir}\,\ast}_{fi}A^{0,{\rm ind}}_{fi}$ 
that are odd with respect to time reversal.)\thinspace\ 
Such contributions do not appear if, as in the case of the 
longitudinal hopping conductivity in Sec.~\ref{SEC:OHEinterferenceL}, 
one considers one-phonon processes~\cite{REF:numphon}.  They do, however, 
appear if one considers, e.g., two-phonon processes described by 
combinations of amplitudes obeying the following property: 
$A^{0,{\rm dir}}_{fi}$ contains an even (odd) number of complex 
energy denominators when $A^{0,{\rm ind}}_{fi}$ contains an odd 
(even) number.  Then these denominators give rise to an additional 
energy-conserving $\delta$-function~\cite{exception}, a quantity 
that is odd respect to time reversal [cf.~Eq.~(\ref{prob})], and 
yields contributions to the probability that behave suitably under 
time-reversal. 
\begin{figure}[hbt]
\epsfxsize=\columnwidth
\centerline{\epsfbox{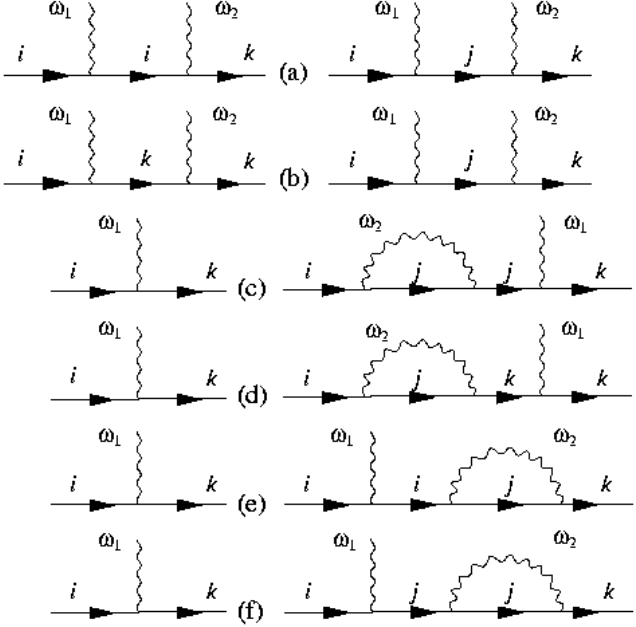}}
\vskip0.4cm
\caption{Pairs of interfering amplitudes that result in the 
Hall effect. 
Rows~(a) and (b) each show the interference of two-stage 
processes. 
Rows~(c) to  (f) each show the interference of a one-stage 
and a three-stage process.  
Right column: indirect hopping processes; 
left column:    direct hopping processes. 
Lines with arrows correspond to carrier propagators; 
their intersections with wavy lines correspond to 
matrix elements of carrier-phonon interactions.}
\label{FIG:set}
\end{figure}%
Having ascertained the structure of the amplitudes that give rise 
to the Hall effect, we now select and examine the dominant 
contributing processes (i.e.~those involving the smallest possible 
number of electron-phonon interactions).  These involve two-phonon 
transitions and, as shown by Holstein~\cite{REF:Holstein1}, their 
elements (i.e.~direct and indirect processes) can be visualized as 
follows: 
\hfil\break\noindent
(i)~Both transitions are two-stage processes (i.e.~involve two 
transitions between exact carrier states.  For example, the 
indirect and indirect transitions respectively being 
\begin{mathletters}
\begin{eqnarray}
(i, N_1, N_2)
&\rightarrow&
(j, N^{\prime}_1, N_2)
\rightarrow
(k, N^{\prime}_1,  N^{\prime}_2), 
\\
(i, N_1, N_1)
&\rightarrow&
(i, N^{\prime}_1, N_2)
\rightarrow  
(k,N^{\prime}_1,  N^{\prime}_2),
\end{eqnarray}
\end{mathletters}%
where 
$N_1$, 
$N^{\prime}_1$,  
$N_2$ and 
$N^{\prime}_2$ are the occupation numbers of phonon modes 1 and 2, 
$N$, and $N^{\prime}$ differ by unity,
$i$, $j$ and $k$ label initial, intermediate and final sites 
(as well as carrier states localized on these sites).
\hfil\break\noindent
(ii)~The direct transition is a one-stage process, e.g.,
\begin{mathletters}
\begin{equation}
(i, N_1)
\rightarrow 
(k, N^{\prime}_1), 
\end{equation}
and the indirect transition is a three-stage process, e.g.,
\begin{eqnarray}
&&(i, N_1, N_2)
\rightarrow
(i, N^{\prime}_1, N_2)
\nonumber
\\
&&\qquad\qquad
\rightarrow
(j, N^{\prime}_1, N^{\prime}_2)
\rightarrow
(k, N^{\prime}_1, N_2). 
\end{eqnarray}
\end{mathletters}%
(The whole set of processes includes those which result from alterations of
the sequences of various subprocesses). One can observe, that 
phonon modes whose population is changed in processes (i) or (ii),
interact with charge carrier on both direct and indirect path, and, as we
shall see, these paths can interfere~\cite{occupation}. 
Complete set of processes that lead to the Hall effect is shown on 
Fig.~\ref{FIG:set}.

\eqbreak

In terms of phonon-assisted transitions between exact 
carrier eigenstates $\ket{\Psi_i}$, the two-stage processes 
are described by the amplitudes 
\begin{mathletters}
\begin{eqnarray}
&&A^{\rm dir,2}_{k, N_1,N_2; j, N^{\prime}_1, N^{\prime}_2 }=
\nonumber\\
\noalign{\medskip}
&&\qquad
\phantom{+}\frac{\me{\Psi_k , N_1}{H_{\rm el-ph}}{\Psi_k , N^{\prime}_1}
\me{\Psi_k , N_2}{H_{\rm el-ph}}{\Psi_i , N^{\prime}_2} -
\me{\Psi_k , N_2}{H_{\rm el-ph}}{\Psi_i , N^{\prime}_2}
\me{\Psi_i , N_1}{H_{\rm el-ph}}{\Psi_i , N^{\prime}_1}}{E_j -E_k}
\nonumber\\
&&\qquad
+\frac{\me{\Psi_k , N_2}{H_{\rm el-ph}}{\Psi_k , N^{\prime}_2}
\me{\Psi_k , N_1}{H_{\rm el-ph}}{\Psi_i , N^{\prime}_1} -
\me{\Psi_k , N_1}{H_{\rm el-ph}}{\Psi_i , N^{\prime}_1}
\me{\Psi_i , N_2}{H_{\rm el-ph}}{\Psi_i , N^{\prime}_2}}{E_i -E_j}
\label{EQ:twostagedir}\\
&&A^{\rm ind,2}_{k, N_1,N_2; j, N^{\prime}_1, N^{\prime}_2 }=
\nonumber\\
\noalign{\medskip}
&&\qquad
\frac{\me{\Psi_k , N_1}{H_{\rm el-ph}}{\Psi_j , N^{\prime}_1}
\me{\Psi_j , N_2}{H_{\rm el-ph}}{\Psi_i , N^{\prime}_2}}{E_i-E_j + 
(N^{\prime}_1 -N_1)\hbar\omega_1 +i\hbar\gamma} +
\frac{\me{\Psi_k , N_2}{H_{\rm el-ph}}{\Psi_j , N^{\prime}_2}
\me{\Psi_j , N_1}{H_{\rm el-ph}}{\Psi_i , N^{\prime}_1}}{E_i-E_j + 
(N^{\prime}_2 -N_2)\hbar\omega_2 +i\hbar\gamma},
\label{EQ:twostageind}
\end{eqnarray}%
\end{mathletters}%
where 
$A^{\rm dir,2}_{k, N_1,N_2; j, N^{\prime}_1, N^{\prime}_2 }$ 
and 
$A^{\rm ind,2}_{k, N_1,N_2; j, N^{\prime}_1, N^{\prime}_2 }$ 
are, respectively, the amplitudes of the two-stage-direct and 
two-stage-indirect processes.  The interfering amplitudes for 
the one-stage process $A^{\rm dir,1}$ and (an example of) a 
three-stage process $A^{\rm ind,3}$ are given in terms of the 
exact carrier eigenstates $\ket{\Psi_i}$:  
\begin{mathletters}
\begin{eqnarray}
&&A^{\rm dir,1}=
\me{\Psi_k , N_1}{H_{\rm el-ph}}{\Psi_i , N^{\prime}_1}, 
\label{EQ:onestage}\\
&&A^{\rm ind,3}=
\frac
{\me{\Psi_k , N_2}{H_{\rm el-ph}}{\Psi_j , N^{\prime}_2}
\me{\Psi_j , N^{\prime}_2}{H_{\rm el-ph}}{\Psi_j , N_2}
\me{\Psi_j , N_1}{H_{\rm el-ph}}{\Psi_i , N^{\prime}_1}}
{(E_i-E_j)(E_i-E_j+(N^{\prime}_1 - N_1)\hbar\omega_1+i\hbar\gamma)}. 
\label{EQ:threestage}
\end{eqnarray}
\end{mathletters}%
\eqresume

Let us briefly discuss the energy conservation (i.e.~$\delta$-function) 
structure and time-reversal properties of the probabilities associated 
with the amplitudes given in 
Eqs.~(\ref{EQ:twostagedir}), (\ref{EQ:twostageind}), (\ref{EQ:onestage}) 
and (\ref{EQ:threestage}).
The explicit $\delta$-function in the formula~(\ref{EQ:transrate})  
for $W_{fi}$ constrains the energies of the initial and final states.  
One further $\delta$-function arises when we insert explicit expressions 
for the interfering pairs of amplitudes, 
Eqs.~(\ref{EQ:twostagedir},\ref{EQ:twostageind}) and
Eqs.~(\ref{EQ:onestage},\ref{EQ:threestage}), 
into Eq.~(\ref{EQ:transrate}. This second  $\delta$-function 
characterizes energy conservation between 
initial and intermediate (or intermediate and final states). 
As we mentioned earlier, time reversal symmetry can also be satisfied 
if we use, e.g.,  imaginary parts of Lorentzian functions instead of 
$\delta$-functions, thus taking into account approximate energy 
conservation in transitions between broadened states of the system.    
Here, for brevity, we use the term $\delta$-function when discussing 
time-reversal properties of transition rates. 

Let us follow how these $\delta$-functions (and, hence, the requisite 
odd character under time reversal) emerge. For two-stage processes, 
a direct transition has to contain one intermediate state, which has to 
be virtual state, and, therefore Eq.~(\ref{EQ:twostagedir}) contains 
real energy denominators. Thus, in two stage 
processes, a $\delta$-function additional to one 
giving energy conservation between initial and final state, 
is to arise from indirect transition amplitudes. Such a 
$\delta$-function
indeed arises (in each of the contributing terms) from complex 
denominators of Eq.~(\ref{EQ:twostageind}), upon summation over all 
possible phonon modes. 
In interference involving three-stage processes,  
the analytical expression for three-stage amplitude  
Eq.~(\ref{EQ:threestage}) 
is characterized by two energy denominators. Upon summation over all 
possible phonon modes, one of these denominators leads to a $\delta$-function. 
Another energy denominator in Eq.~(\ref{EQ:threestage}) 
corresponds to a virtual transition.
As follows from Eq.~(\ref{EQ:sinphi}),
in the presence of the Aharonov-Bohm phase 
$\phi$ picked up by carriers moving around
three sites, interference contributions to 
$W_{if}$ determined by Eqs.~(\ref{EQ:twostagedir},\ref{EQ:twostageind}) and by
Eqs.~(\ref{EQ:onestage},\ref{EQ:threestage}) all contain two 
$\delta$-functions, and, as required by time-reversal symmetry 
properties, are proportional to $\sin{\phi}$. 

In order to see the physical meaning of interference between 
amplitudes of direct and indirect two-stage processes or 
interference between amplitudes of one- and three-stage processes, 
it is instructive 
to write down the amplitudes~(\ref{EQ:twostagedir}), 
(\ref{EQ:twostageind}), (\ref{EQ:onestage}) and (\ref{EQ:threestage}) 
in terms of the 
ionic orbitals $\ket{\phi_i}$ and the transfer integrals $V_{ij}$, 
by using Eq.~(\ref{expansion}).
(The energy denominators are corrected compared to 
those that can be found in the original Holstein paper~\cite{REF:Holstein1}).
This will also allow us to formulate conditions necessary for coherence 
of two-phonon processes, relevant for the Hall effect.

\eqbreak

In terms of  $\ket{\phi_i}$ and  $V_{ij}$, the two-stage processes 
are described by the amplitudes 
\begin{mathletters}
\begin{eqnarray}
&&A^{\rm dir,2}_{k, N_1,N_2; j, N^{\prime}_1, N^{\prime}_2 }=
\nonumber\\
\noalign{\medskip}
&&\qquad
\phantom{+}
\frac{\me{\phi_k , N_1}{H_{\rm el-ph}}{\phi_k , N^{\prime}_1}
\me{\phi_i , N_2}{H_{\rm el-ph}}{\phi_i , N^{\prime}_2}V_{ki} -
\me{\phi_k , N_2}{H_{\rm el-ph}}{\phi_k , N^{\prime}_2}
\me{\phi_i , N_1}{H_{\rm el-ph}}{\phi_i , N^{\prime}_1}V_{ki}}
{(E_j -E_k)(E_i-E_k)}
\nonumber\\
&&\qquad
+\frac{
\me{\phi_k , N_2}{H_{\rm el-ph}}{\phi_k , N^{\prime}_2}
\me{\phi_i , N_1}{H_{\rm el-ph}}{\phi_i , N^{\prime}_1}V_{ik} -
\me{\phi_k , N_1}{H_{\rm el-ph}}{\phi_k , N^{\prime}_1}
\me{\phi_i , N_2}{H_{\rm el-ph}}{\phi_i , N^{\prime}_2}V_{ik}}
{(E_i -E_j)(E_k-E_i)}
\label{EQ:phitwostagedir}\\
&&A^{\rm ind,2}_{k, N_1,N_2; j, N^{\prime}_1, N^{\prime}_2 }=
\nonumber\\
\noalign{\medskip}
&&\qquad
\frac{\me{\phi_k , N_1}{H_{\rm el-ph}}{\phi_k , N^{\prime}_1}
\me{\phi_i , N_2}{H_{\rm el-ph}}{\phi_i , N^{\prime}_2}V_{jk}}
{(E_i-E_j + (N^{\prime}_1 -N_1)\hbar\omega_1 +i\hbar\gamma)(E_k-E_j)} +
\frac{\me{\phi_k , N_2}{H_{\rm el-ph}}{\phi_j , N^{\prime}_2}
\me{\phi_j , N_1}{H_{\rm el-ph}}{\phi_i , N^{\prime}_1}}
{E_i-E_j + (N^{\prime}_2 -N_2)\hbar\omega_2 +i\hbar\gamma}.
\label{EQ:phitwostageind}
\end{eqnarray}%
\end{mathletters}%
The interfering amplitudes for the one-stage process $A^{\rm dir,1}$ 
and an example of a three-stage process $A^{\rm ind,3}$ which, 
in terms of exact carrier states, are given by 
Eqs.~(\ref{EQ:onestage} and (\ref{EQ:threestage})
in terms of ionic orbitals read:  
\begin{mathletters}
\begin{eqnarray}
&&A^{\rm dir,1}=
\frac
{\me{\phi_k , N_1}{H_{\rm el-ph}}{\phi_k , N^{\prime}_1}V_{ki}+
\me{\phi_i , N_1}{H_{\rm el-ph}}{\phi_i , N^{\prime}_1}V_{ki}}
{E_i-E_k}, 
\label{EQ:phionestage}\\
&&A^{\rm ind,3}=\cdots +
\frac
{\me{\phi_j , N_2}{H_{\rm el-ph}}{\phi_j , N^{\prime}_2}
\me{\phi_j , N^{\prime}_2}{H_{\rm el-ph}}{\phi_j , N_2}
\me{\phi_i , N_1}{H_{\rm el-ph}}{\phi_i , N^{\prime}_1}V_{kj}V_{ji}}
{(E_j-E_k)(E_i-E_j)
(E_i-E_j)(E_i-E_j+(N^{\prime}_1 - N_1)\hbar\omega_1+i\hbar\gamma)}. 
\label{EQ:phithreestage}
\end{eqnarray}
\end{mathletters}%

\eqresume

First term in Eq.~(\ref{EQ:phionestage}) does not contribute 
to interference of one- and three-stage processes, because 
the matrix element $\me{\phi_k , N_1}{H_{\rm el-ph}}{\phi_k , N^{\prime}_1}$ 
does not correspond to any time-reversed counterpart in 
Eq.~(\ref{EQ:phithreestage}). We are now in a position to discuss 
the physical meaning of processes that contribute to the Hall effect 
in terms of local orbitals. In the interference of two stage processes,
both amplitudes, direct and indirect, correspond to interaction with 
a phonon mode  $(N_1, \omega_1)$ at site $k$ (initial state $\ket{\phi_k}$, 
tunneling to site $i$ 
(final state $\ket{\phi_i}$ ) directly or via intermediate 
site $j$, and interacting with 
a phonon mode $(N_2, \omega_2)$ (at site $i$) 
that is distinct from one participating in a 
process occurring at site $k$. In the interference of one- and three-stage 
processes, the direct one-stage process (which is, strictly speaking, 
can be called one-stage only in terms of exact carrier states) can include 
two possibilities: (i) 
Interaction with a phonon mode $(N_1, \omega_1)$ 
at initial site and tunneling to the 
final site; (ii) tunneling to the final site and interaction with 
a phonon mode $(N_1, \omega_1)$ at the final site. Then the three 
stage process have to include the following stages: 
In the case (i) there is interaction with the phonon mode 
$(N_1, \omega_1)$ at initial site, tunneling to intermediate site 
$j$, emission (absorption) and reabsorption (reemission) 
of a phonon mode $(N_2, \omega_2)$ at site $j$, and tunneling
to the final site. In the case (ii) charge carrier tunnels to 
intermediate site $j$, where emission (absorption) and 
reabsorption (reemission) of a phonon mode $(N_2, \omega_2)$   
occurs, and then tunnels to the final site.     
We therefore see that, in terms of local orbitals, processes 
contributing to the Hall effect are characterized by interference 
of amplitudes in which phonon modes changing their occupation numbers 
are represented by time reversed counterparts in $A^{\rm dir}$ and 
$A^{\rm ind}$, respectively. 

Having described the inelastic processes leading to the Hall effect,
we are now in the position to generalize conditions for occurrence of 
interference, and to  formulate  
these conditions for two-phonon processes. It follows from 
Eqs. (\ref{EQ:twostageind},\ref{EQ:twostagedir},\ref{EQ:onestage},
\ref{EQ:threestage}) that  
electron-phonon interaction results in coherence of transfer amplitudes  
in two cases:
(i)~If direct transition and transition via an intermediate state 
both occur as two-phonon processes~\cite{REF:numphon}, 
with two phonon modes changing their 
occupation numbers in the course of both these transitions, then interference 
exists when
the two phonons leading to the direct transition are the same as two phonons 
leading to a transition via intermediate site;
(ii)~If the  
direct transition occurs as one-stage process assisted by a
phonon mode, this phonon mode changes its occupation number, and
the transition involving an intermediate site occurs as a 
three stage process. 
Then, the condition is that one of the phonon modes assisting three-stage 
process is the same as that in the direct transition, while another
phonon mode which assists the indirect transition does not 
change its occupation number. 

Broadly speaking, both these cases lead to conditions that 
inelastic modes that change their state in the course of electronic hop
are the same in the two interfering hopping amplitudes. Only in this case an 
interference exists even between amplitudes of
inelastic processes. We note that this point has been also recently revisited 
by Entin-Wohlman et al.~\cite{REF:Entin}.

\subsubsection{Ordinary Hall effect: 
Local conductivities. Remarks on averaging over triads}
\label{SEC:OHEaot}

These interference contributions will result
in the ordinary Hall effect, with the Hall 
conductivity in a triad of ions $\sigma_{\rm OH}$ given by 
\begin{equation}
\sigma_{\rm OH}=
G\{\epsilon_{j}\}\,
\sin\big({\bf B}\cdot{\bf\arQ}/\phQ\big),
\label{EQ:hallcond}
\end{equation}
where $\phQ$ is the (electromagnetic) flux quantum, 
${\bf\arQ}$ is the (oriented, real space) area of the triad, and 
$\{\epsilon_{j}\}_{j=1}^{3}$ are the energies of the three single-particle 
eigenstates, which are invariant under reversal of the AB flux.  
The explicit expression for $G$ can be found by substituting 
Eqs.~(\ref{EQ:phitwostagedir}), (\ref{EQ:phitwostageind}), 
(\ref{EQ:phionestage}) and (\ref{EQ:phithreestage})  
into Eqs.~(\ref{EQ:transrate}) and (\ref{EQ:hallcur}) .  
(Note that in generic case $G$ also depends on 
the populations of these states, which 
themselves may depend on particle-particle correlations.)\thinspace\ 

In~\cite{REF:Holstein1}, Holstein mainly addressed the issue of the  
Hall effect in hopping conductors in the presence of an a.c.~electric 
field.  Compared to the d.c.~Hall effect, the a.c.~problem is simplified 
because the principal contribution to the Hall conductance comes from 
those spatially isolated configurations of sites for which the population 
relaxation time $t_{\rm r}$ is on the order of the inverse frequency 
$\omega$ of the current (i.e.~$\omega t_{\rm r}\sim 1$). 
In this a.c.~case, there is no need to address the (highly nontrivial) 
issue of how these sources of the Hall effect (i.e.~configurations of 
sites) are combined into a conducting network that is connected to the 
Hall contacts.  For the d.c.~Hall effect, on the other hand, this issue 
of the structure of the conducting network must be faced, and it becomes 
necessary to understand which triads are the most effective in 
contributing to the Hall effect, how to average over triads, and what 
quantity should be averaged.  [If the quantity to be averaged should be 
the conductivity (resistivity) then one should first compute the local 
conductivities (resistivities) of triads and then obtain the macroscopic 
conductivities (resistivities) by averaging.]\thinspace\  This issue of 
averaging over all triads and conducting network structure in disordered 
systems remains controversial~\cite{REF:Galperin,REF:Entin}.  

However, in the present Paper we are concerned not with the ordinary 
Hall effect in a system with localized carriers but the anomalous one.  
For the Anomalous Hall effect, the Aharonov-Bohm phase does not play a 
fundamental role, a very weak magnetic field being applied solely for 
the purpose of inducing a macroscopic magnetization of a ferromagnetic 
medium.  Rather, for the AHE in system comprising magnetically 
disordered core-spins on Mn sites visited by hopping charge carriers, 
it is a certain type of {\it spin\/} quantal 
phase~\cite{REF:LeeNaga,REF:Wiegmann,REF:Dagotto} 
that manifests itself.  We now turn the origin and meaning of this  
spin quantal phase.  

\subsection{The quantal Pancharatnam phase}
\label{SEC:qp}
\begin{figure}[hbt]
\epsfxsize=3.0in
\centerline{\epsfbox{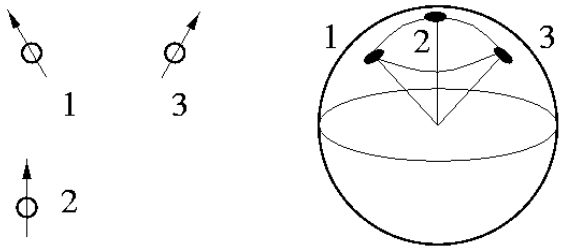}}
\vskip+0.20truecm
\caption{(a)~Triad of Mn ions having distinct core spin orientations. 
         (b)~Sphere of possible core spin orientations, showing  
         the specific orientations of the spins for this triad.  
         The orientations form the vertices of spherical triangle. 
         The area of this triangle determines the quantal Pancharatnam 
         phase.}
\label{FIG:triad}
\end{figure}%
To understand the nature of the spin quantal phases, let us begin 
by examining the single-particle quantum mechanics of a carrier 
hole added to a triad of ${\rm Mn}^{3+}$ ions.  We regard the 
spin-3/2 core spins of the Mn ions as large enough to be treated 
classically, so that one can assign a definite direction to each 
of them.  Thus, a generic configuration of core spins is 
characterized by the triad of unit vectors 
$\big\{
\spv_{1},
\spv_{2},
\spv_{3}
\big\}$, 
respectively located at the triad of sites 
$\big\{
\pov_{1},
\pov_{2},
\pov_{3}
\big\}$, 
as depicted in Fig.~\ref{FIG:triad}. Let us now consider the transfer 
of the hole carrier between ions.  In the double exchange model, such 
transfer is described by the Hamiltonian 
\begin{eqnarray}
H_{\rm DE} &=&
\sum
\limits_{\alpha,j}
\ket{\alpha,\phi_j}\epsilon_j\bra{\alpha,\phi_j} 
+\sum\limits_{{\alpha\atop{j\ne k}}}
\ket{\alpha,\phi_j}V_{jk}\bra{\alpha,\phi_k}
\nonumber\\
&&\qquad\qquad
+J\sum\limits_{{\alpha,\beta\atop{j}}}
\ket{\alpha,\phi_j}\,\spv_{j}
\cdot\mbox{\boldmath $\sigma$}_{\alpha,\beta}\,\bra{\beta,\phi_j}, 
\label{EQ:de}
\end{eqnarray}  
where $\mbox{\boldmath $\sigma$}$ is the Pauli spin operator describing 
the spin of the hole carrier, $J$ is the Hund Rules coupling energy 
which, for ${\rm Mn}^{3+}$ ions, is on the order of several eV, and is 
much larger than the orbital transfer integrals $V_{jk}$, and 
$\ket{\alpha,\phi_j}$ is the outer-shell atomic carrier state at site 
$j$, the index $\alpha$ labeling the spin-projection on to the $z$ axis 
(i.e.~$\ket{\alpha}\vert_{\alpha=\pm}$ are eigenstates of $\sigma_{z}$).
In general, the Hamiltonian~(\ref{EQ:de}) results in spin polarization of 
electrons at arbitrary ratio of $J$ and  $V_{jk}$. However, two limiting 
cases are of interest. At $J\ll \vert{V_{jk}}\vert$, metallic 
ferromagnetism is usually treated perturbatively in terms of an electron 
gas in metals, resulting in the RKKY interaction. In this case charge carriers 
with any spin projection are taken into account. The opposite case 
$J\gg \vert{V_{jk}}\vert$ is a purely double-exchange model. This is the 
case that is relevant to manganites.
By Hund's Rules (appropriate for $J\gg \vert{V_{jk}}\vert$), there is, 
at each site, effectively a single quantum state available to the hole.  
This state $\ket{\spv_j,\phi_j}$ is the one in which the carrier spin 
projection opposes the core spin direction.  The orbital $\ket{\phi_j}$ 
that characterizes this state is {\it one\/} of the orbitals of an 
isolated Mn ion, centered on the site $j$; we choose to omit the 
remaining orbitals so as to simplify the discussion.  As for the 
other spin state (as well as any other orbitals), we regard them as 
being inaccessible on energetic grounds.  Postponing until later any 
effects of spin-orbit interactions, we assume that the transfer of 
carriers (being, as it is, effected by either the \lq\lq kinetic 
energy\rlap,\rq\rq\ by the second term in Eq.~(\ref{EQ:de}), or by the 
electron-phonon interaction) has no effect on the spin of the 
carriers.  The hopping amplitudes between ionic states are thus 
given by 
$\me{\spv_k,\phi_k}{T}{\spv_j,\phi_j}$, 
where $T$ corresponds either to the transfer operator $V$ of 
Eq.~(\ref{EQ:el}) or to the electron phonon interaction $H_{\rm el-ph}$. 
These hopping amplitudes depend explicitly on the relative orientation 
of the core spins, $\spv_k$ and $\spv_j$.  In particular, by projecting 
the Hamiltonian~(\ref{EQ:de}) on to the physically relevant low-energy 
subspace spanned by the states $\ket{\spv_j,\phi_j}$, we arrive at the 
the double-exchange Hamiltonian projected on to the low-energy 
subspace, $H^{\prime}_{\rm DE}$, which takes the form
\begin{mathletters} 
\begin{eqnarray} 
&&H^{\prime}_{\rm DE}
=
\sum\limits_j \ket{\spv_j,\phi_j}
(\epsilon_j -J)
\bra{\spv_j,\phi_j} 
\nonumber\\
&&
\qquad\qquad\qquad+
\sum\limits_{j\ne k}
\ket{\spv_j,\phi_j}
{V^{\rm eff}_{jk}}
\bra{\spv_k,\phi_k}, 
\label{eff}
\\
&&
V_{jk}^{\rm ef}
\equiv
\me{\spv_j, \phi_j}{H_{\rm DE}}{\spv_k, \phi_k}
\nonumber\\
&&\quad
=V_{jk}\left(
\cos{\frac{\theta_j}{2}}\cos{\frac{\theta_k}{2}}+
e^{i\gamma_{kj}}
\sin{\frac{\theta_j}{2}}\sin{\frac{\theta_k}{2}}
\right),
\label{EQ:effha}
\end{eqnarray}
\end{mathletters} 
where $V_{jk}^{\rm ef}$ are the effective transfer amplitudes, 
$\gamma_{kj}\equiv\gamma_k-\gamma_j$ and, respectively, 
$\theta_j$ and $\theta_k$ are the azimuthal angles and 
$\gamma_j$ and $\gamma_k$ are the polar angles 
of the semiclassical spin directions $\spv_j$ and $\spv_k$. 
Provided we choose, e.g., 
$\spv_j\parallel{\bf e}_{z}$ and  
$\spv_k\parallel{\bf e}_{x}$ 
(where $\{{\bf e}_{x},{\bf e}_{y},{\bf e}_{z}\}$ 
are the Cartesian basis vectors) this effective transfer amplitude 
reduces to the Anderson-Hasegawa form: 
\begin{equation}
V_{jk}^{\rm AH}=V_{jk}\cos{\theta/2},
\label{EQ:effAH}
\end{equation} 
where $\theta$ is the angle between $\spv_j$ and $\spv_k$. 
However, and this is central to our discussion, the effective 
transfer amplitude is, in general, a complex quantity characterized 
by its amplitude and phase. 
From Eqs.~(\ref{EQ:effha} and (\ref{EQ:effAH}), it is 
indeed apparent that if core spins are co-aligned, the effective hopping
amplitude is maximal, while if the core spins on two ions are opposite, 
hopping between such ions does not occur.

We now follow the line of argument applied in 
Sec.~\ref{SEC:OHEinterferenceL}, and 
construct the exact eigenstates of Hamilitonian~(\ref{eff}), via 
Eq.~(\ref{expansion}) modified to account for spin.  Hence, we can 
build matrix elements of the electron-phonon interaction between exact 
localized states, taking into account the effect of the core spin 
orientations $\{\spv_{j}\}$.  We now note that this expansion for the 
state gives rise, in the hopping probability $\vert{U_{ji}}\vert^{2}$ 
(and hence in the hopping rate $W_{ji}$), to terms containing products 
of matrix elements such as transfer integrals $V^{\rm eff}_{kj}$ and 
electron-phonon interactions $U_{kj}$.  Amongst these terms are ones 
containing matrix elements associated with paths around closed 
loops and incorporating the effects of interference between distinct 
carrier paths.  The simplest example involves the product 
$V^{\rm eff}_{ij}V^{\rm eff}_{kj}V^{\rm eff}_{ji}$, 
associated with the path $i\to j\to k\to i$.

In the presence of the constraints set by the core spin orientations, 
the transfer of carriers discussed in the previous paragraph is subject 
to a striking quantal effect.  To see this, consider the products of 
matrix elements $V^{\rm eff}_{ik}V^{\rm eff}_{kj}V^{\rm eff}_{ji}$. 
Explicitly, such products have the form
\begin{eqnarray}
&&
V^{\rm eff}_{ik}V^{\rm eff}_{kj}V^{\rm eff}_{ji}
\nonumber\\
&&
\qquad=
\me{\spv_i,\phi_i}{H_{\rm DE}}{\spv_k,\phi_k}
\nonumber\\
&&\qquad\qquad\times
\me{\spv_k,\phi_k}{H_{\rm DE}}{\spv_j,\phi_j}
\me{\spv_j,\phi_j}{H_{\rm DE}}{\spv_i,\phi_i}
\nonumber\\
&&
\qquad=
\bra{\spv_i}\otimes\me{\phi_i}{H_{\rm DE}^{\prime}}{\phi_k}\otimes\ket{\spv_k}
\nonumber\\
&&\qquad\qquad\times
\bra{\spv_k}\otimes\me{\phi_k}{H_{\rm DE}^{\prime}}{\phi_j}\otimes\ket{\spv_j}
\nonumber\\
&&\qquad\qquad\times
\bra{\spv_j}\otimes\me{\phi_j}{H_{\rm DE}^{\prime}}{\phi_i}\otimes\ket{\spv_i}
\nonumber
\\
&&
\qquad\propto
\ipr{\spv_i}{\spv_k}
\ipr{\spv_k}{\spv_j}
\ipr{\spv_j}{\spv_i}
={\rm Tr}\,P_k\,P_j\,P_i, 
\end{eqnarray}
where the operators
$P_j\equiv(1+\mbox{\boldmath$\sigma$}\cdot{\bf n}_j)/2$ are 
projectors (in spin space) on to the spin states aligned with 
the local core spin orientations $\spv_{j}$.  From this last 
expression, in terms of projectors, it is straightforward to 
establish that 
\begin{eqnarray}
&&
{\rm Tr}\,P_k\,P_j\,P_i 
\\
&&
\quad= 
\big(1+
\spv_{1}\cdot\spv_{2}+
\spv_{2}\cdot\spv_{3}+
\spv_{3}\cdot\spv_{1}\big)
+i({\spv_{1}\cdot(\spv_{2}\times\spv_{3})}).
\nonumber
\end{eqnarray}
Hence, we arrive at the quantal phase $\Omega$, the phase 
of the complex quantity ${\rm Tr}\,P_k\,P_j\,P_i$, 
which is given by 
\begin{equation}
\frac{\Omega}{2}=
\tan^{-1}\frac{\spv_{1}\cdot(\spv_{2}\times\spv_{3})}
{1+\spv_{1}\cdot\spv_{2}+\spv_{2}\cdot\spv_{3}+\spv_{3}\cdot\spv_{1}}.
\label{EQ:PanchSolid}
\end{equation}
In the context of the physical quantity from which the computation 
of the quantal phase $\Omega$ emerged, viz., the perturbative evaluation of
the hopping rate $W_{kj}$, Eq.~(\ref{EQ:transrate}),  
this phase modulates the interference between hopping processes that 
progress from one site $j$ to another site $k$, either directly or 
indirectly, via a third site.  

Formula Eq.~(\ref{EQ:PanchSolid}) indicates that the phase $\Omega$ 
has a geometric interpretation as the (oriented) solid angle of the 
geodesic triangle having vertices at $\{\spv_i,\spv_j,\spv_k\}$ on 
the unit sphere.  It is the quantal analog of the classical optical 
phase discovered in the context of polarized light by 
Pancharatnam~\cite{REF:Pancha,REF:BonPan}.  In that setting, what 
Pancharatnam showed is that a under a sequence of changes of the 
polarization state of light that return the light to its original 
polarization state there arises a phase shift (i.e.~a phase anholonomy) 
determined by the geometry of the sequence of changes.  If the 
sequence of polarization states is represented by a sequence of points 
on the Poincar\'e sphere (a certain parameterization of light 
polarizations) then $\Omega$ is given by the area of the geodesic 
polygon on this sphere the vertices of which are correspond to these 
polarization states.

In the double-exchange electronic analog of Pancharatnam's phase, the 
transporting of an outer-shell carrier to an ion with a differently 
oriented core spin via a spin-independent process is characterized 
by a matrix element that can be interpreted as a {\it connection\/}. 
For processes visiting a closed sequence of sites, this connection 
yields a quantal phase $\Omega$, viz., the phase shift of the returning 
spin state in terms of the sequence of sites visited.  This quantal 
version of Pancharatnam's phase is given by {\it half\/} the area of 
the geodesic polygon (on the unit sphere of core spin orientations) 
the vertices of which are the core spin orientations of the sites visited. 
Although the phase $\Omega$ emerged from considerations of interference 
between processes involving a {\it triad\/} of sites, such phases are 
more general and would, in fact, emerge for arbitrary processes. 
We remark that, in contrast to Berry's adiabatic phase~\cite{footnote}, 
the phenomenon described here is associated with {\it sudden\/} changes 
in the carrier-spin state, and need not be slow.

\subsection{The Anomalous Hall effect in hopping regime}
\label{SEC:hophall}

We now turn from the OHE in a spinless triad to the  AHE in a triad of 
magnetic sites.  Like the OHE given by Eq.~\ref{EQ:hallcond}, 
this AHE results from two-phonon processes, 
but is due to the Pancharatnam phase instead of the AB phase.  (At this 
stage, we have not yet included the effects of the spin-orbit 
interaction.)\thinspace\ {\it Mutatis mutandis\/}, we arrive at the AH 
conductivity $\sigma_{\rm AH}$, given by 
\begin{equation}
\sigma_{\rm AH}=
  G\{\varepsilon_{j}\}\,
  \cos\frac{\theta_{13}}{2}\,
  \cos\frac{\theta_{32}}{2}\,
  \cos\frac{\theta_{21}}{2}\, 
  \sin\frac{\Omega}{2},
\label{EQ:AHcurrent}
\end{equation}
where $\cos\theta_{jk}\equiv\spv_{j}\cdot\spv_k$, the factors 
$\cos(\theta_{jk}/2)$ are Anderson-Has{\-}egawa factors, and 
$\{\varepsilon_{j}\}$ are the energies of the three on-site 
single-particle eigenstates that are consistent with Hund's Rules, 
these energies depending on 
$\{\spv_{j}\cdot\spv_{k}\}_{1\le j<k\le 3}$ and $\cos(\Omega/2)$.  
Note that $G$ is even under the reversal of the Pancharatnam flux 
$\Omega\to-\Omega$, and $\sigma_{\rm AH}$ is odd under it. 

\begin{figure}[hbt]
\epsfxsize=3.0in
\centerline{\epsfbox{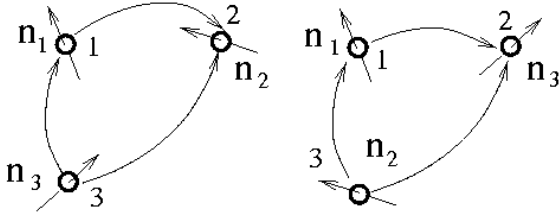}}
\vskip+0.20truecm
\caption{Two configurations [(a) and (b)] of core spins on a triad of 
Mn ions, differing by the interchange of the spins on sites 2 and 3.  
These distinct spin configurations give rise to opposite Pancharatnam 
phases.}
\label{FIG:compen}
\end{figure}%

We have shown that, for a triad with given set of core-spin orientations, 
an AHE arises from the quantal Pancharatnam flux.  However, there is a 
significant difference between this AHE and the OHE.  In the former 
(nonmagnetic) case, a uniform applied magnetic field leads to a net 
macroscopic OHE, even though contributions of triads may cancel one 
another~\cite{REF:Galperin}.  In the latter case (with its magnetic sites, 
Pancharatnam flux, but spin-orbit interactions not yet included), even 
the presence of a macroscopic magnetization of the core spins is insufficient 
to cause a {\it macroscopic\/} Hall current.  The reason for this is that 
in obtaining the macroscopic AH current from Eq.~(\ref{EQ:AHcurrent}) we 
must average over the configurations of the core spins.  In the absence of 
spin-orbit interactions, the distribution of these configurations, although 
favoring a preferred {\it direction\/} (i.e., the magnetization direction 
${\bf m}\equiv{\bf M}/M$), is invariant under a reflection of all 
core-spin vectors in any plane containing the magnetization, and, therefore, 
there is no preferred Pancharatnam flux. For example, two spin configurations
shown in Fig.~\ref{FIG:compen} have the same magnetization but opposite signs
of the Pancharatnam flux. This fact, 
coupled with the fact that $\{\varepsilon_{j}\}$ are also invariant 
under such reflections, guarantees that the macroscopic AH current will 
average to zero.  (We do, however, expect significant AH current 
{\it noise\/} in the ferromagnet-paramagnet transition regime, owing to 
the fluctuations of the Pontryagin charge~\cite{REF:IoLeMi} of the triads 
of core spins (which we shall define shortly) and, hence, elementary 
Pancharatnam fluxes.) 

In order to capture the AHE in double-exchange materials such as 
manganites, we must consider not only the Pancharatnam phase but 
also some agent capable of lifting the reflection invariance of the 
energies $\{\varepsilon_{j}\}$ and the distribution of core-spin 
configurations and, hence, capable of inducing sensitivity to the 
sign of the Pancharatnam flux. Such an agent is provided by 
spin-orbit interactions.  We now discuss the effect of these 
interactions on the motion of charge carrier in a triad. 

\subsection{Spin-orbit interactions in a triad}
\label{SEC:soitr}

The most general form of spin-orbit interaction is given by 
the spin-orbit Hamiltonian 
\begin{equation}
H_{\rm so}=\alpha{\bf p}\cdot
\left(\mbox{\boldmath $\sigma$}\times
\mbox{\boldmath $\nabla$}U\right), 
\label{EQ:soi}
\end{equation}
where the potential $U$ includes ionic and impurity potentials, $\alpha$ 
is the spin-orbit interaction constant, ${\bf p}$ is the electron momentum, 
and $\mbox{\boldmath$\sigma$}$ are the Pauli operators.  This spin-orbit 
interaction results in an effective SU(2) gauge potential 
${\bf A}_{\rm so}=\alpha m (\mbox{\boldmath $\sigma$}
\times\mbox{\boldmath $\nabla$}U)$~\cite{Goldhaber}, 
where $m$ is the relevant mass of the carrier.  This gauge potential 
provides an additional source of quantal phase.  For a given core-spin 
configuration, the spin-orbit interaction favors one sense of 
carrier-circulation around the triad over the other, and thus favors 
one sign of Pancharatnam phase over the other. 

Let us consider the consequences for the energy spectrum of the triad 
$\{\varepsilon_{j}\}$ that arise due to spin-orbit interactions. 
This interaction generates a dependence of 
$\{\varepsilon_{j}\}$ on the three vector-products 
$\Spv_{jk}\equiv\spv_{j}\times\spv_{k}$ 
which, together with the magnetization direction ${\bf m}$, yield a 
preferred value for the triad Pontryagin charge 
$\pcd$ [$\equiv\spv_{1}\cdot(\spv_{2}\times\spv_{3})$] and, hence, 
a preferred Pancharatnam flux. 

It is straightforward to find corrections, due to the SOI, of hole 
eigenenergies if the on-site energies of the holes are nondegenerate. 
Then the sensitivity of $\{\varepsilon_{j}\}$ to vector-products 
$\Spv_{jk}\equiv\spv_{j}\times\spv_{k}$ first enters 
at third order (in the transfer matrix elements):
\begin{equation}
\delta\varepsilon_j\!=\!\sum_{h,k(\ne j)} 
{\rm Tr}\,T_{jh}\,T_{hk}\,T_{kj}\big/
(\varepsilon_j\!-\!\varepsilon_h)(\varepsilon_j\!-\!\varepsilon_k),
\end{equation} 
where $T_{jk}\equiv P_j\,V_{jk}\,P_k$ are the transfer amplitudes, 
$V_{jk}$ are the hopping matrix elements, and 
${\rm Tr}$ denotes a trace in spin space. (For degenerate on-site 
hole energies 
one should obtain the splitting of these energies due to transfer in 
the absence of SOI, and then include SOI at the final step, arriving at 
the result to be given below.)\thinspace\  The hopping matrix elements 
are sensitive to the SOI quantal phase, and can be written in the form 
$V_{jk}= V_{jk}^{\rm orb}\,L_{jk}$, where 
$L_{jk}\equiv\big(1+i\mbox{\boldmath$\sigma$}\cdot{\bf g}_{jk})$, 
$V_{jk}^{\rm orb}$ is an orbital factor, and ${\bf g}_{jk}$ 
($\propto\alpha_{\rm so}$) is an appropriate vector that describes 
the average SOI for the transition $j\rightarrow k$ in a triad of sites 
$i$, $j$ and $k$.  
Then, e.g., the first-order (in $\alpha$) shifts in the 
$\varepsilon$'s are given by 
\begin{eqnarray}
&&
\delta\varepsilon_j\propto 
{\rm Tr}\,T_{13}\,T_{32}\,T_{21}=
4\,{\rm Re}\,{\rm Tr}\,
P_1\,L_{13}\,P_3\,L_{32}\,P_2\,L_{21}
\nonumber\\
&&\quad=
2\left(
\Spv_{13}\cdot\sov_{13}+
\Spv_{32}\cdot\sov_{32}+
\Spv_{21}\cdot\sov_{21}\right)-\Spv\cdot\sov,
\label{EQ:correction}
\end{eqnarray}%
where $\Spv\equiv\Spv_{13}+\Spv_{32}+\Spv_{21}$, 
and   $\sov\equiv\sov_{13}+\sov_{32}+\sov_{21}$. 
If the potential $U$ in the SOI is a superposition of 
spherically-symmetric ionic potentials in a the triad of sites 
then the vectors 
$\sov_{jk}$ have a transparent geometrical meaning: 
\begin{eqnarray}
\sov_{jk}&=&a_{jk}{\bf\arQ}, 
\\
{\bf\arQ}&=&\frac{1}{2}
\left(\pov_j-\pov_h\right)
\times
\left(\pov_k-\pov_h\right), 
\label{EQ:VecArea}
\end{eqnarray}  
i.e., they are proportional to the area $\vert{\bf\arQ}\vert$ 
of the triangle whose vertices are the sites $\pov_j$, $\pov_k$ 
and $\pov_h$.  
Then the SOI-generated shift in the carrier eigenenergies 
has the Dzyaloshinski-Moriya form~\cite{REF:Dzyaloshinski}.

\subsection{Elementary Hall conductivity in a triad}
\label{SEC:elemhall}

There are two contributions to the AHE which result from 
the SOI-generated shift in the carrier eigenenergies.
The first contribution is due to the dependence of the 
probability of hopping around the triad on $\{\varepsilon_{j}\}$ 
for a given spin configuration.  
By incorporating the shifts~(\ref{EQ:correction}), together with 
the Pancharatnam phase, we arrive at the elementary AH conductivity 
\begin{equation}
\sigma^{(1)}_{\rm AH}=  
\spv_{1}\cdot(\spv_{2}\times\spv_{3})
\sum\nolimits_{j}\delta\varepsilon_{j}\,
\partial G/\partial\varepsilon_{j}.
\label{EQ:AHEC}
\end{equation}
As discussed above, Eq.~(\ref{EQ:AHEC}) 
has a nonzero macroscopic average, owing to the presence of a 
characteristic Pontryagin charge constructible from the $\Spv_{jk}$, 
that feature in the energy shifts, and the magnetization direction. 
A second consequence of the SOI-generated carrier-energy 
shift~(\ref{EQ:correction}) leads to the second contribution,  
$\sigma^{(2)}_{\rm AH}$.  Due to the feedback of the (fast) carrier 
freedoms, which provide an effective potential for the (slow) spin 
system, determined by Eq.~(\ref{EQ:correction}), the 
equilibrium probabilities of spin configurations having opposing 
Pancharatnam fluxes will no longer be equal.  (For this contribution, 
which is related not to $\partial G/\partial\varepsilon_{j}$ but to 
$G$ itself, there is no need to account for SOI-induced carrier-energy 
shifts in the current now being averaged over a nonsymmetric 
spin-configuration distribution.)\thinspace\  A contribution with
this origin has also been considered in Ref.~\cite{Ye}.  
$\sigma^{(1)}_{\rm AH}$ and $\sigma^{(2)}_{\rm AH}$ are of the 
same order of magnitude. 

\subsection{Structure of the conducting network and the AHE resistivity}
\label{SEC:structure}

We now consider the question of how the physics of elementary triads 
of Mn ions relates to the macroscopic properties of manganites.  For 
hopping conductivity, the pathways taken by the current depends 
sensitively on the details of the configuration of the core spins, 
owing to the sensitivity of the hopping amplitudes to the core-spin 
alignments.  In particular, regions having local spin configurations 
that are aligned roughly opposite to the macroscopic magnetization of 
the sample tend to be avoided by the current.  This fact renders rather 
subtle the procedure for averaging over equilibrium spin configurations, 
which must account for effects such as local spin correlations and 
excitations of various types (i.e.~spin wave and topological 
excitations).   

En route to computing the AH resistivity, let us try to identify 
through which triads the AH current tends to flow.  Carriers tend to 
pass through regions of lower resistance.  However, currents through 
regions with aligned spins do not lead to an AHE because the relevant 
Pancharatnam flux through such regions is small. 
\begin{figure}[hbt]
\epsfxsize=\columnwidth
\centerline{\epsfbox{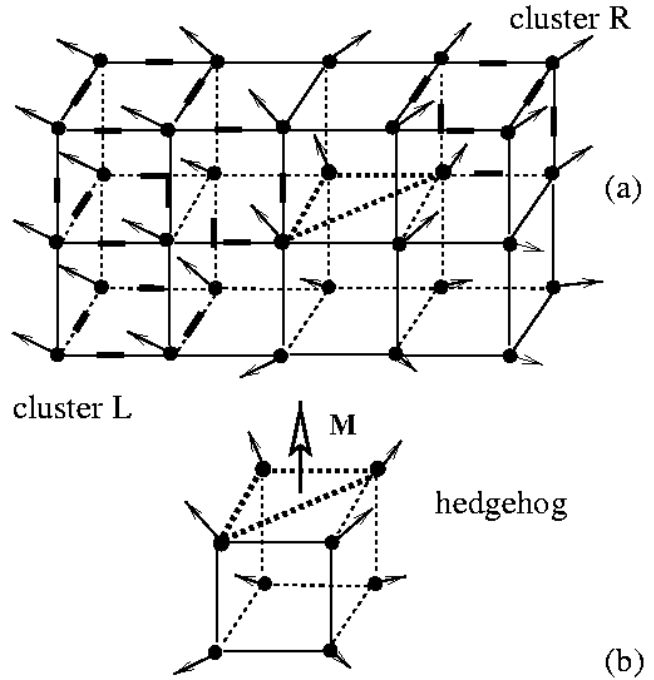}}
 \vskip0.3cm
\caption{(a)~Two clusters of sites, denoted $L$ and $R$, 
in a fragment of the sublattice of Mn sites. 
Within a cluster the core spin orientations are roughly the same; but 
spins in distinct clusters have significantly different orientations. 
Note that the boundary between such clusters is liable to contain 
spin configurations that resemble the single hedgehog configuration 
shown in~(b).  The heavy dotted lines in~(a) indicate a triad of spins 
that contribute to a hedgehog configuration.  Sites in the upper part 
of the cell shown in~(b) contribute to the conducting network and, 
correspondingly, the magnetization (per site) in this upper part of the 
cell (shown as an open-headed arrow) is roughly that of the sample (see 
the discussion in this section).}
\label{FIG:struct}
\end{figure}
\noindent
 Furthermore, in the ferromagnet-to-paramagnet transition region, 
where the magnetization 
is only a small fraction of its saturation value, even those spins 
within the network responsible for longitudinal conductivity have 
orientations that are typically splayed, relative to one another.   
Nevertheless, if one were to consider only the spins in this network, 
their average magnetization would be larger than that of the entire 
sample, and the typical solid angles formed by triads of such spins  
would be rather small.
Thus, any AHE originating in such triads 
would be not the dominant contribution.  Moreover, close to the 
metal-insulator transition, hopping paths through the sample that 
encounter spins having rather common orientations do not exist.
In fact, there is experimental evidence for the existence of moderately 
sized clusters of aligned spins coming from neutron scattering 
data~\cite{Lynn,neutron} at temperatures near the ferromagnet-to-paramagnet 
transition.  These data indicates that clusters of spins do indeed exist 
in which spins are aligned over a length scale of roughly $10\,\AA$.  
Thus, it is reasonable to envision the magnetic configurations as 
comprising rather well oriented clusters of, say, twenty to thirty 
spins, with adjacent clusters having rather different spin orientations.
Furthermore, theoretical estimates of the scale of magnetic fields 
relevant for colossal magnetoresistance are consistent with the 
existence of such clusters. (Such clusters can be regarded as being 
large magnetic polarons~\cite{Pokrovsky}.)\thinspace\  
As the characteristic Zeeman temperature associated with a {\it single\/} 
spin-3/2 in a magnetic field of $7\,{\rm T}$ is $20\,{\rm K}$, whereas 
the characteristic temperature associated with colossal magnetoresistance 
is roughly $200\,{\rm K}$, one is lead to the view that clusters of 
correlated spins involve on the order of ten spins.

Now consider two adjacent clusters of roughly aligned spins 
(e.g.~clusters $L$ and $R$ in Fig.~\ref{FIG:struct}a.  Even the 
conducting paths connecting these clusters contain bonds between ions 
having significantly misaligned spins.  These spins belong to regions 
of magnetic inhomogeneity, e.g., inhabited by hedgehog excitations, 
(an example of \lq\lq lattice hedgehog\rq\rq\ 
is shown in Fig.~\ref{FIG:struct}b) which defines the border 
of aligned clusters as shown in Fig.~\ref{FIG:struct}a.  
As mentioned in the 
Introduction, hedgehog excitations are long-lived topological 
spin excitations, the existence of which is known to be important 
for ferromagnet-to-paramagnet transition~\cite{Lau,Murthy}, even in 
the three dimensional case.  

Within these regions of magnetic inhomogeneity there are triads of 
splayed spins.  Let us now address the question: What is the 
characteristic splay?  To answer this question, imagine dividing 
up the spins into those within clusters and those within the 
border regions.  Even though the magnetization per spin in a 
typical cluster is greater than the sample-average magnetization 
per spin, the clusters are misoriented relative to one another. 
Thus, the contribution to the magnetization per spin of the sample 
coming from the spins in clusters is not guaranteed to exceed the 
sample average and, indeed, it seems reasonable to assume that it 
is, in fact, not so different from the sample average.  If so, then 
the magnetization per spin of the spins in the border region would 
also be roughly the sample average.  We shall make the hypothesis that 
this is indeed the case.  Then the magnetization of typical triads of 
ions in the border region can also roughly be taken to be the average 
sample magnetization.  We shall denote by $\beta$ the characteristic 
angle that spins in a triad form with the direction of the overall 
magnetization of the sample ${\bf M}$: 
$\cos\beta\sim\vert{\bf M}\vert/M_{\rm s}$, 
where $M_{\rm s}$ is the saturation magnetization of the sample.

Let us now imagine how charge carriers move between clusters having 
quite different ion spin orientations.  Such motion is necessary for 
the existence charge-carrier propagation between electrical contacts. 
We have sketched a typical instantaneous configuration of the spins 
in Fig.~\ref{FIG:struct}.   As one can see, the \lq\lq upper\rq\rq\ 
part of a inhomogeneous region formed on a cubic lattice of sites 
can serve as a path for hopping between clusters.  For reasons that 
we will now give, triads of sites within this (and similar) regions 
are effective contributors to the AHE: 
\hfil\break\noindent
(i)~The three spins in the triad have positive components along 
the direction of magnetization.  (Recall that clusters with 
magnetization pointing against the majority tend to be avoided by 
the current.)\thinspace\  This allows participation by these 
triad sites in the conducting network.  If all three sites participate 
in the conducting network, the triad can be an effective source of an  
electromagnetic force that leads to a Hall effect. 
As we have discussed, the net magnetization of the triad is 
roughly that of the bulk, which makes it magnetically compatible 
with its neighbors.      
\hfil\break\noindent
(ii)~Typical triads of spins, being located as they are between 
several misoriented clusters are significantly splayed.  Therefore, 
the solid angle formed by their spins (i.e.~the area of the geodesic 
triangle formed by their orientations on the unit sphere, also known 
as the Pontryagin charge of the spin configuration) is substantial and, 
in fact, close to the maximum possible value given the constraint 
that the triad magnetization (per spin) be comparable to the sample 
magnetization (per spin).  Thus, we adopt as a caricature of the 
spin configuration in regions contributing to the Hall effect a 
picture of splayed triads of spins of known magnetization density, 
residing within tetrads of spins on a lattice plaquette, such as 
those depicted by the lattice hedgehog configuration shown in 
Fig.~\ref{FIG:struct}. 
This scheme, in which we consider tetrads of a given magnetization 
and then select triads of sites in a tetrad, seems to us appropriate, 
given the cubic structure of the sublattice of Mn spins.  However, 
alternative schemes (e.g.~in which one considers triads themselves 
or other assemblies of splayed spins of a given magnetization rather than 
tetrads, and chooses triads out of these assemblies) lead to 
almost identical results (e.g.~for the scaling 
of the Hall resistivity, which we discuss in the present and 
following sections). 
This insensitivity to details is all the more natural, given that 
we are dealing with an atomically disordered system.
We note that, because hedgehogs are topologically 
stable, they provide a mechanism by which the spin configuration 
can sustain strongly splayed regions that persist for durations 
much longer than the characteristic time for charge motion.  Hence, 
in their presence, on can accurately treat the charge motion as taking 
place with a background of inhomogeneous but essentially static spins, 
which renders consistent the adiabatic treatment of the dynamics of 
the spins.

Thus, we arrive at the notion of an {\it optimal triad\/}. 
An optimal triad is a triad of spins residing in a tetrad 
of four spins around a plaquette of the cubic sublattice of 
Mn ions and having the following properties: 
(i)~The tetrad has the magnetization density of the bulk; and 
(ii)~subject to this constraint, the spins of the tetrad are 
maximally splayed (i.e.~subtend the maximal solid angle and, in 
fact, are configured symmetrically around a cone).  
 Note that if the lattice were triangular then we 
would simply have adopted a definition of optimality in 
terms of maximally splayed triads (rather than tetrads) of 
spins.  As mentioned above, in disordered 
systems (such as manganites), the distinctions engendered by such 
options are unlikely to have a strong impact on the physical 
consequences of the picture. 

The motion of charge carriers through optimal triads gives rise to the AHE. 
We note that these optimal triads have properties quite different 
from those of optimal triads contributing to the OHE in doped, 
non-magnetic semiconductors~\cite{REF:Galperin}: in the OHE setting, 
only two sites in an optimal OHE triad are connected to the 
conducting network, whereas in the optimal AHE triads all three 
triad sites participate in the network. (Indeed, if alternatively, 
one of the sites is {\it not\/} a part of the conducting network, 
its spin  must be roughly opposite that of the spins on the other 
two sites; such a configuration would yield only a small 
Pancharatnam phase.) 

The question may arise why triads within tetrads (and not, for 
instance, tetrads themselves) are considered to be the dominant 
source of the AHE.  By a contribution from a tetrad we mean one 
involving four overlap integrals.  As these overlap integrals are 
small, owing to the localized character of the carrier wavefunctions 
and, thus, the contribution from tetrads is suppressed, relative to 
that from triads.  We note that distortions due to doping, 
particularly deviations of Mn-O-Mn bond angles from 180 degrees, 
facilitates tunneling between Mn ions via plaquette diagonals  
(see Fig.~\ref{FIG:struct}). As was estimated in~\cite{Jaime-PRL}, 
the amplitude of transfer along diagonals is 0.5 of that between nearest 
neighbor Mn ions. Recent tight binding model parameterization 
of local density approximation (LDA) studies~\cite{REF:Satpathy} show 
that hopping via diagonals is even more important, and its amplitude 
is 0.82 of transfer amplitude  between nearest neighbor Mn ions.   

Having discussed the structure of resistive network, let us now 
calculate the longitudinal and Hall resistivities of manganites 
in the regime in which conductivity proceeds by hopping (i.e.~at 
temperatures above, as well as somewhat below, the 
ferromagnet-to-paramagnet transition). 
The longitudinal hopping conductivity arising from phonon-assisted 
hops between sites $i$ and $j$ is given by 
\begin{equation}
\sigma _{xx}=(ne^2d^2/k_{\rm B}T)W_0^{ij}\cos^2(\theta /2),
\label{hopcond}
\end{equation}
[c.f.~Eqs.~(\ref{EQ:current}) and (\ref{EQ:Ohm})],
where $d$ is the distance between sites.  Here, $W_0^{ij}$ is the 
rate of phonon-assisted direct hops, and we have explicitly 
separated out the Anderson-Hasegawa factor $\cos^2(\theta /2)$. 
Correspondingly, the (anomalous) Hall conductivity is given by 
[c.f.~Eq.(~\ref{EQ:hallcur})]
\begin{equation}
\sigma_{xy}=(ne^2d^2/k_BT)W_1^{ij},
\label{hopAHcon}
\end{equation}
where $W_1^{ij}$ is the rate of hopping between the two sites, and
accounts for interference associated with both direct hopping 
and hopping via an intermediate state on a third site. Note that the
quantity $W_1$ includes three Anderson-Hasegawa factors 
[and so does the Hall conductivity given by Eq.~(\ref{EQ:AHcurrent})]. 
The task of computing the Hall resistivity $\rho_{xy}$, which 
in the limit of $\sigma_{xx}\gg\sigma_{xy}$ under consideration 
has the form
\begin{equation}
\rho_{xy}\approx-\frac{\sigma_{xy}}{\sigma^2_{xx}}, 
\label{Hallres}
\end{equation}
then reduces to a determination of a ratio involving the direct 
and indirect hopping rates $W_{0}^{ij}$ and $W_{1}^{ij}$
(as a function of the magnetization texture). 
As discussed in Secs.~\ref{SEC:holst} and \ref{SEC:hophall}, 
$W_1^{ij}$ involves two-phonon processes, 
whereas $W_0^{ij}$ involves only single-phonon processes.
Because of this, dependence on electron-phonon coupling constant, 
phonon occupation numbers, and charge carrier 
occupation numbers, cancels from the relevant 
ratio, $W_1^{ij}/(W^{ij}_0)^2$, so that this ratio can be written as
\begin{equation}
W_1/W_0^2=\alpha\hbar\zeta/k_{\rm B}T,
\label{EQ:ratio}
\end{equation}
where $\alpha$ is a numerical factor describing the multiplicity of 
the various carrier-phonon interference processes 
(see REf.~\cite{REF:Holstein1} and  Sec.~\ref{SEC:holst}), the number of 
intermediate sites, and the difference between nearest- and 
next-nearest-neighbor hopping amplitudes.  We shall refer to the 
parameter $\zeta$, which characterizes the difference between 
the forward  $W_1^{ij}$ (backward $W_1^{ji}$) transition rates, 
as an asymmetry parameter.  For the OHE, this asymmetry parameter 
is given by 
\begin{equation}
\zeta\propto\sin({\bf B}\cdot{\bf Q}/\phi _0),
\end{equation}
where ${\bf Q}$ is the vector area defined in 
Eq.~(\ref{EQ:VecArea}), as follows 
from Eq.~(\ref{EQ:hallcur}).

For the AHE, it follows from Eqs.(~\ref{EQ:correction}) 
and (\ref{EQ:AHEC}) that the assymmetry parameter is given by 
\begin{equation}
\zeta
\simeq 3[{\bf g}_{jk}\cdot ({\bf n}_j\times 
{\bf n}_k)][{\bf n}_1\cdot ({\bf n}_2\times {\bf n}_3)]/4, 
\end{equation}
where ${\bf g}_{jk}$ are characteristic vectors
arising in the hopping amplitude owing to the spin-orbit 
quantal phase; 
${\bf n}_j$ are unit vectors of the core spins in the triad, 
and ${\bf n}_1\cdot ({\bf n}_2\times {\bf n}_3)$ 
is the volume of a parallelepiped defined by
core-spin vectors, i.e., the Pontryagin charge $q_{\rm P}$. 
The anomalous Hall resistivity can
be written in the simple form 
\begin{equation}
\rho _{xy}\simeq 
-\sigma _{xy}/\sigma _{xx}^2=
-\frac 1{ne}
\left( 
\frac{\alpha \hbar \zeta }{ed^2}
\frac 1{\cos ^4(\theta /2)}
\right) .  
\label{eq2}
\end{equation}
The evaluation of Eq.~(\ref{eq2}) reduces to a determination of 
$\theta$, along with the products $({\bf n}_j\times {\bf n}_k)$ and 
${\bf n}_1\cdot ({\bf n}_2\times {\bf n}_3)$, which survive averaging 
over all possible triads.  The dominant contribution to the average of 
these products arises from optimal spin  
configurations.
\begin{figure}[hbt]
\epsfxsize=8cm
\centerline{\epsfbox{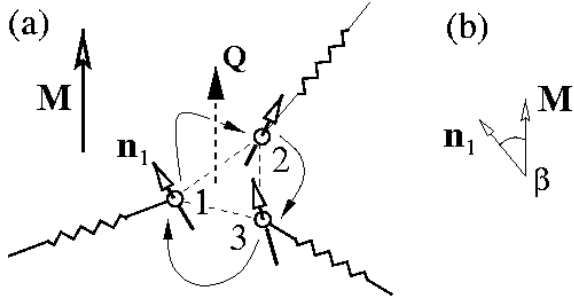}}
\vskip0.5truecm
\caption{(a)~Triad of Mn sites (1, 2 and 3) in the conducting network.  
Charge-carrier motion around triads such as this lead to the AHE.  
We compute the longitudinal and AH conductivities by relating them 
to the configuration of the spins (which we suppose to be optimal) 
and, hence, to the magnetization.}
\label{FIG:aver}
\end{figure}%

Therefore, in line with properties~(i) and (ii) of these 
configurations, consider a square lattice formed by the Mn 
ions in a plane perpendicular to ${\bf m}$ (as in, e.g., 
the top surface of the cube in Fig.~\ref{FIG:struct}b).  To 
ascertain the geometry of optimal triads, consider the spins at 
four sites of a plaquette belonging to the conducting network.  
Being optimally configured, these spins lie at equal separations 
around a cone whose vertical angle $2\beta$ is given by 
$2\cos ^{-1}[M(H,T)/M_{\text{sat}}]$.  (The angle between the 
altitude of the cone and any generator in the conical surace is 
$\beta$.)\thinspace\  Let us now use this information 
to fix the various geometrical quantities that determine 
the longitudinal and Hall conductivities, the former being 
associated with pairs of sites and the latter with triads.
We assert that to compute the contributions to these conductivities 
we may consider sites in an optimal configuration, as these are 
characteristic of that regions of the sample that dominantly contribute.  
For the Hall conductivity, the reasons for accepting this assertion 
were discussed in the present subsection.  As for the longitudinal 
resistivity, the assertion is valid because this quantity is dominated 
by the most resistive regions of the conducting network, and these are 
expected to arise at the interface between clusters of aligned spins, 
i.e., in regions that are hedgehog-like.  

With this picture in mind, we now compute characteristic values for 
the geometrical quantities that feature in the longitudinal and Hall 
conductivities.  Thus, we need 
the angle $\theta$ between adjacent spins, 
the Pontryagin charge $q_{\rm P}$, and the 
products $({\bf n}_j\times {\bf n}_k)$, 
each of which is related to the vertical 
angle $2\beta$ by elementary geometry: 
\begin{mathletters} 
\begin{eqnarray} 
2\cos^2({\theta/2)}
&=& 1+\cos^2\beta,                      \\  
q_P&=&2\cos \beta \,\sin^2\beta,        \\ 
{\bf m}\cdot({\bf n}_j\times{\bf n}_k)
&=&\sin^2\beta. 
\end{eqnarray} 
\end{mathletters} 

We now have to account for the fact that, in the hopping regime, 
the magnitude of the longitudinal 
(and anomalous Hall) resistivities 
depend on the probability that pairs and triads of ions 
are connected to the conducting network. 
We introduce a
percolation factor $P$ describing the connectivity of the
pair to the conducting network; 
for the AH conductivity the corresponding 
factor would be $%
P^2$ because both pairs of ions in a triad must, as discussed above, 
belong to the
conducting network. It is remarkable that, throughout the 
localization regime, $\rho _{xy}$
is nevertheless determined by currents formed in individual pairs and
triads, because the factors of $P$ cancel in the expression for 
$\rho _{xy}$ given by Eqs.~(\ref{Hallres}, \ref{eq2}). Therefore, in so far 
as $q_{\rm P}$ and
the angles between neighboring spins can be directly related to 
$m\equiv M/M_{\text{sat}}=\cos \beta $, the Hall resistivity 
$\rho _{xy}$ depends on $H$ and $T$ only
through $m(H,T),$ and is given by. 
\begin{equation}
\rho _{xy}=\rho _{xy}^0\frac{m(1-m^2)^2}{(1+m^2)^2}.  
\label{eq3}
\end{equation}

To determine the magnitude of the AHE, we first need to estimate
the characteristic values of $|{\bf g}_{jk}|\sim g$ arising from the
spin-orbit interaction (SOI). As discussed in Sec.~\ref{SEC:soitr}, 
the SOI term leads to a
Dzyaloshinski-Moriya contribution to the eigenenergy of the 
carriers. A standard estimate~\cite{REF:Landau} gives 
the characteristic values of 
$\vert {\bf g}\vert\sim g\sim Ze^2/4m_ec^2d_0$, where $d_0$ is the radius of
an Mn core d-state. 
 While renormalization of carrier parameters in
crystals may tend to increase $|{\bf g}_{jk}|,$ crystalline symmetry 
requires admixtures of core orbitals, which, in turn, are mixed with 
oxygen p-orbitals, with outer-shell wavefunctions 
in order to have $|{\bf g}_{jk}|\neq 0$ . Such admixture 
is effectively generated by the non-collinearity of the
Mn-O-Mn bonds that allows carrier hopping around triads\ (including jumps
along plaquette diagonals).An estimate based on free electron parameters gives 
$g\sim 5\times 10^{-4}$. (We note that 
band-structure calculation of the spin-orbit coupling constants is 
outside the scope of this paper).
The characteristic strength of the
Dzyaloshinski-Moriya terms is  $\sim gt_0 \sim 0.02$ meV, and the 
characteristic strength of the spin-orbit interaction is 
$\sim \epsilon t_0 \sim 0.1$ meV, where $\epsilon$ is the characteristic
carrier energy. Not only these strengthss are much smaller 
than the characteristic double 
exchange energy, but they are even smaller than the magnitude of the direct 
antiferromagnetic Heisenberg exchange term. However, for the anomalous 
Hall effect in the localized regime, the Dzyaloshinski-Moriya 
terms are crucial, as we discussed in Sec.~\ref{SEC:hophall}.

We now estimate the macroscopic longitudinal and
Hall resistivities in the regime in which the 
conducting network is fully connected, i.e., in the regime 
IV of Fig.~\ref{FIG:tran}. 
By taking $n=5.6\times 10^{21}$ cm$^{-3}$, $W_0\sim
2.5\times 10^{13}$ s$^{-1}$, and, from the magnetization
data at $T=275$ K (Fig. 1),  $\cos \beta =0.6$, 
we obtain $\rho _{xx}\simeq 1$ m$\Omega $ cm
which coincides with the value of the experimentally observed resistivity
for LPMO (see Fig.~\ref{FIG:longrho}). 
The AHE contribution to the Hall resistivity, 
assuming a numerical
factor $\alpha$ of 2.5, is then $\rho _{xy}\approx -0.5$ $\mu \Omega $ cm, in
agreement with the experimentally observed LPMO 
Hall resistivity at the same $T$
(Fig.~\ref{FIG:hall2}). 
The equivalent expression for the hopping Hall resistance in the
Holstein mechanism is defined by the asymmetry parameter 
$\zeta \simeq \cos ^2(\theta /2)\cos \beta \sin ({\bf %
B}\cdot {\bf Q}/\phi _0)$ and, at $B=1$ T, is an order of magnitude smaller
than the AHE. We expect the macroscopic hopping AH and OH effects 
to have the
same sign, opposite to that of the OHE in the metallic regime.

In the next section, Sec.~\ref{SEC:disctheorex}, 
we shall compare the results for the Hall resistivity 
with the experimental data. As we shall see, the picture for  
the core spin configurations developed above,
which include clusters of oriented spins and hedgehog-configured spins, 
allows us to explain not only the AHE, but ferromagnet-to-paramagnet and 
metal-insulator transitions in manganites, and provide 
a quantitative explanation of 
the magnitude of characteristic 
magnetic field that result in colossal magnetoresistance.  
The notion of an optimal triad enables us to
fit the experimental data 
for the AHE to a functional dependence of the resistivity on magnetization 
given by Eq.~(\ref{eq3}). The agreement 
of the hopping picture and experimental data 
in the transitional region is remarkable.
 
As we have mentioned above, the structure of the conducting network 
leading to the ordinary Hall effect in disordered doped semiconductors and
the averaging procedures in these systems are still
controversial~\cite{REF:Entin,REF:Galperin}. In contrast to disordered doped
semiconductors, manganites turn out to be systems in
which the ability to tune average magnetization allows one 
to tune optimal triads, whose solid angles (the Pancharatnam phases) 
determine the AHE. The magnetization in manganites, 
therefore, serves as a scaling variable that has 
no analog in OHE in nonmagnetic disordered systems, and 
provides a check on our understanding of the conduction network.  
We note that the presence or absence of small polarons in the system 
does not change the scaling of the AHE resistivity, because, as 
follows from studies of polaronic transport 
in~\cite{REF:Holstein,REF:Friedman,REF:Emin,REF:Entin}, 
Eq.~\ref{EQ:ratio} also holds for small polaron hopping.  

\section{Hall resistivity: comparison of theory and experiment}
\label{SEC:disctheorex}

\begin{figure}
\vskip0.5cm
\epsfig{figure=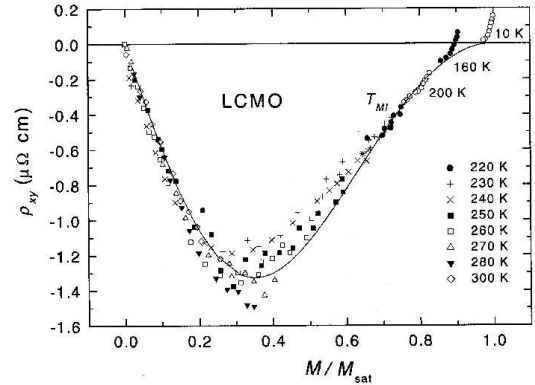,width=3.0in,rheight=3.5in,angle=90,silent=}    
\vskip-3.0cm
\caption{Hall resistivity $\rho_{xy}$ of LCMO versus reduced 
magnetization $M/M_{\rm sat}$ for the data shown in Fig.~4.  
Note the scaling behavior, i.e., the extent to which $\rho_{xy}$ can be 
regarded as depending on $T$ and $B$ solely through $M/M_{\rm sat}$. 
The solid line is a fit to Eq.~(3.52) 
with $\rho_{xy}^0$ = -6.2 $\mu \Omega $cm.}
\label{FIG:sclc}
\end{figure}

\begin{figure}
\vskip0.2cm
\epsfig{figure=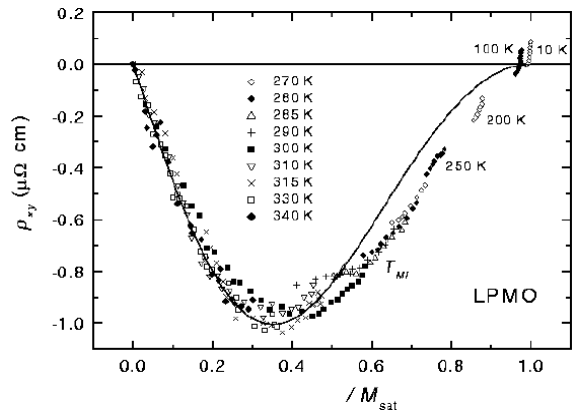,width=3.0in,rheight=3.5in,angle=-90,silent=}    
\vskip-3.0cm
\caption{
Hall resistivity $\rho_{xy}$ of LPMO versus reduced 
magnetization $M/M_{\rm sat}$ for the data shown in Fig.~5.  
Note the scaling behavior, i.e., the extent to which $\rho_{xy}$ can be 
regarded as depending on $T$ and $B$ solely through $M/M_{\rm sat}$. 
The solid line is a fit to Eq.~(3.52) 
with $\rho_{xy}^0$ = -4.7 $\mu \Omega $cm.}
\label{FIG:sclp}
\end{figure}

\begin{figure}
\vskip0.2cm
\epsfig{figure=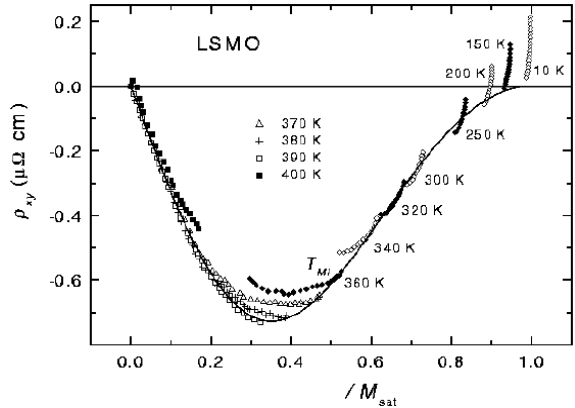,width=3.0in,rheight=3.5in,angle=-90,silent=}    
\vskip-3.0cm
\caption{
Hall resistivity $\rho_{xy}$ of LSMO versus reduced 
magnetization $M/M_{\rm sat}$ for the data shown in Fig.~6.  
Note the scaling behavior, i.e., the extent to which $\rho_{xy}$ can be 
regarded as depending on $T$ and $B$ solely through $M/M_{\rm sat}$. 
The solid line is a fit to Eq.~(3.52) 
with $\rho_{xy}^0$ = -3.4 $\mu \Omega $cm.}
\label{FIG:scls}
\end{figure}

The scaling of the Hall resistivity is shown in 
Figs.~\ref{FIG:sclc}, \ref{FIG:sclp} and \ref{FIG:scls}, 
in which the data shown in Figs.~\ref{FIG:hall1}, \ref{FIG:hall2} 
and \ref{FIG:hall3}, respectively, are replotted as a function of 
$M/M_{\text{sat}}$.  At and above $T_c$ the data fall on a smooth 
curve that reaches an extremum at 
$M/M_{\text{sat}}\simeq 0.4$ for LSMO and LPMO and at 
$M/M_{\text{sat}}\simeq 0.35$ for LCMO. 
Below $T_c$ the data first change rapidly with
magnetization as domains are swept from the sample before saturating and
following the general trend. At the lowest temperatures, the metallic OHE
appears as a positive contribution at constant magnetization. As for 
the magnitude of the Hall resistivity,
\noindent
for LPMO the solid
curve in Fig.~\ref{FIG:sclp} 
follows Eq.~(\ref{eq3}) with $\rho _{xy}^0=-4.7$ $\mu \Omega 
$ cm is consistent with the estimates of $\rho _{xx}$ and $\rho _{xy}$ given
above.
\noindent
 Down to 285 K, which is the Curie temperature determined by
scaling analysis, Eq.~(\ref{eq2}) describes the data for LPMO reasonably well.
In addition, the extremum found from this equation is located at 
$M/M_{\text{sat}}=\cos \beta \approx 0.35$, close
to the experimental extremum. 

In LCMO and LSMO, the agreement between 
theoretical and experimental results is good as well. We note that 
below $T_c,$ the longitudinal resistivity is
metallic and no longer dominated by magnetic disorder. We have not expected 
an agreement between theory and experiment in this range of temperatures, 
but in LCMO and LSMO scaling persists 
at temperatures below $T_{C}$, with notable exception of the range of 
low temperatures and magnetizations close to saturation value, where 
the ordinary Hall effect manifests itself. We note in this regard that 
below $T_{C}$, local spin arrangements still can still dominate the 
AHE via asymmetric scattering or side jumps. 
\noindent
The numerator of Eq.(\ref{eq3}), $m(1-m^2)^2,$
which is characteristic for the behavior of $\sigma _{xy}$ alone, 
has an extremum at $m=1/\sqrt{5}\approx 0.45$ as shown by the dashed 
line in Fig.~\ref{FIG:sclp}.

The broader maximum in
the data suggest a shift toward a hopping model for 
$\rho _{xx}$ and $\rho_{xy}$ 
as the sample is warmed through the metal-insulator transition.

\section{Conclusions}
\label{SEC:concl}

Our investigation of the Hall resistivity, the longitudinal
resistivity, and the magnetization in single crystals of three different 
manganite compounds suggests that near and somewhat above the 
ferromagnet-to-paramagnet transition temperature, transport properties 
are determined by charge carrier hopping between localized states. 
We find both theoretically and experimentally that the Hall resistivity is 
solely determined by the sample magnetization ($M$) near 
and somewhat above the transition temperature. A microscopic model 
for the anomalous Hall effect based on the Holstein picture of the 
ordinary Hall effect in the hopping regime has been proposed and 
explains the results quite well. The 
anomalous Hall effect arises due to interference between 
direct hopping between two sites and hopping via a third site, 
with the quantal phase provided by topologically nontrivial 
configurations of Mn ion core spins in the presence of strong
Hund's rule coupling. These force the hopping charge carrier to follow the
local spin texture, with the average quantal phase arising due to 
local Pancharatnam phases and Dzyaloshinski-Moriya spin-orbit 
interactions. Below the transition temperature, the AHE competes with
the OHE as long-range magnetic order and, presumably, an infinite
percolating metallic cluster, develops.

\section{Acknowledgement}

We are grateful to I.~L.~Aleiner, D.~P.~Arovas, S.~L.~Cooper, E.~Dagotto,
S.~Fishman, J. Lynn, V.~L.~Pokrovsky, H. Roder and  P.~B.~Wiegmann
for useful discussions.
This material is based upon work supported by the U.S. Department of
Energy, Division of Materials Sciences under Award No.~DEFG02-96ER45439
through the University of Illinois Materials Research Laboratory.       
Additional support from NSF-DMR99-75187 is also gratefully acknowledged.

\end{multicols}

\begin{references}
\bibitem[*]{bylineYLG}
E-mail: lyandage@uiuc.edu
\bibitem[\dag]{bylineSHC}
E-mail: 
schun@psu.edu.
Present address: 
Department of Physics, Pennsylvania
State University, University Park, PA 16802-6300.
\bibitem[\ddag]{bylineMBS}
E-mail: salamon@uiuc.edu
\bibitem[\S]{bylinePMG}
E-mail: goldbart@uiuc.edu
\bibitem{Jin}
S. Jin, T. H. Tiefel, M. McCormack, R. A. Fastnacht, R.
Ramesh, and J. H. Chen, Science {\bf 264}, 413 (1994).
\bibitem{Ramirez}
A. P. Ramirez, J. Phys.: Condens. Matter {\bf 9}, 8171 (1997).
\bibitem{Snyder}  
G. J. Snyder, M. R. Beasley, 
T. H. Geballe, R. Hiskes, and
S. DiCarolis, 
Appl. Phys. Lett. {\bf 69}, 4254 (1996); 
G. J. Snyder, R. Hiskes, S. DiCarolis, 
M. R. Beasley, and T. H. Geballe, Phys. Rev. B {\bf 53}, 14 434 (1996).
\bibitem{Jaime-PRL}
M. Jaime, H. T. Hardner, M. B. Salamon, M. Rubinstein,
P. Dorsey, and E. Emin, Phys. Rev. Lett. {\bf 78}, 951 (1997).
\bibitem{Wagner}
P. Wagner, D. Mazilu, L. Trappeniers, V. Moshchalkov,
and Y. Bruynseraede, Phys. Rev. B {\bf 55}, R14 721 (1997).
\bibitem{Matl}
P. Matl, N. P. Ong, Y. F. Yan, Y. Q. Li, D. Studebaker, T.
Baum, and G. Doubinina, Phys. Rev. B {\bf 57}, 10 248 (1998).
\bibitem{Jakob}
G. Jakob, F. Martin, W. Westerburg, and H. Adrian, Phys.
Rev. B {\bf 57}, 10 252 (1998); G. Jakob, W. Westerburg, F. Martin, H.
Adrian, P. J. M. van Bentum, and J. A. A. J. Perenboom, 
J. Appl. Phys. {\bf 85}, 4803 (1999); 
W. Westerburg, F. Martin, G. Jakob, P. J. M. van Bentum,
and J. A. A. J. Perenboom, preprint (cond-mat/9907346).
\bibitem{Asamitsu}
A. Asamitsu and Y. Tokura, Phys. Rev. B {\bf 58}, 47 (1998).
\bibitem{LPMO-PRB}
S. H. Chun, M. B. Salamon, and P. D. Han, 
Phys. Rev. B {\bf 59}, 11,155 (1999), 
J. Appl. Phys. {\bf 85}, 5573 (1999).
\bibitem{LPMO-PRL}
S. H. Chun, M. B. Salamon, P. D. Han, Y. Lyanda-Geller,
and P. M. Goldbart, 
Phys. Rev. Lett {\bf 84}, 757 (2000).
\bibitem{Yuli}
Y. Lyanda-Geller, P. M. Goldbart, S. H. Chun, and M. B. Salamon, 
cond-mat/9904331.
\bibitem{LCMOxover}
S. H. Chun, M. B. Salamon, Y. Tomioka, and Y. Tokura,
Phys. Rev. B {\bf 61}, R9225 (2000).
\bibitem{Mott}
See, e.g. N. F. Mott and H. S. W. Massey, 
{\sl Theory of Atomic Collisions\/} (Oxford, 1933).
\bibitem{Smit}
J. Smit, Physica {\bf 23}, 39 (1958). 
\bibitem{Luttinger}
J. Luttinger, 
Phys. Rev. {\bf 112\/}, 739 (1958). 
\bibitem{Maranzana}
F. E. Maranzana, 
Phys. Rev. {\bf 160}, 421 (1967).
\bibitem{Berger} 
L. Berger, Phys. Rev. B {\bf 2}, 4559 (1970). 
\bibitem{Nozieres}
P. Nozi\`{e}res and C. Lewiner, 
J. Phys. (Paris) {\bf 34\/}, 901 (1973).
\bibitem{YuliZ} 
Y. Lyanda-Geller, Pis'ma v ZHETF, {\bf 46} 388 (1987)
(JETP Letters  {\bf 46} 489 (1987)).
\bibitem{Zener}  
C. Zener, Phys. Rev. B {\bf 82\/}, 403 (1951).
\bibitem{Anderson} P. W. Anderson
and H. Hasegawa, Phys. Rev. {\bf 100\/}, 675 (1955).
\bibitem{DeGennes} P. G. De Gennes, Phys. Rev. B. {\bf 118}, 141 (1960).
\bibitem{REF:Holstein1}  
T. Holstein, Phys. Rev. {\bf 124\/}, 1329 (1961).
\bibitem{Aharonov} 
Y. Aharonov and D. Bohm, 
Phys. Rev. {\bf 122} (1959).
\bibitem{REF:Pancha}
S. Pancharatnam, 
Proc. Ind. Ac. Sc. A~{\bf 44\/}, 247 (1956).
\bibitem{REF:BonPan}  
M. V. Berry, 
Int. J. Mod. Optics {\bf 34\/}, 1401 (1987).
\bibitem{REF:Dzyaloshinski}  
I. E. Dzyaloshinski, J. Phys. Chem. Solids 
{\bf 4\/}, 241 (1958); 
T. Moriya, Phys. Rev. Lett. {\bf 4\/}, 5 (1960). 
For band-like electrons terms of similar type were 
discussed in 
P. M. Levy and A. Fert, 
Phys. Rev. B {\bf 23\/}, 4667 (1981).
\bibitem{REF:Kim} 
Y. B. Kim, P. Majumdar, A. J. Millis and B. I. Shraiman, 
cond-mat/9803350.
\bibitem{Ye}  J. Ye, Y. B. Kim, A. J. Millis, B. I. Shraiman, P. Majumdar,
and Z. Tesanovic, Phys. Rev. Lett., {\bf 83} 3737 (1999). 
\bibitem{REF:YLG} Y. Lyanda-Geller, P. M. Goldbart and I. L. Aleiner,
in preparation.
\bibitem{Millis1}  A. J. Millis, P. B. Littlewood, and B. I. Shraiman, Phys.
Rev. Lett. {\bf 74\/}, 5144 (1995).
\bibitem{Varma} 
C. M. Varma, 
Phys. Rev. B {\bf 54\/}, 7328 (1996).
\bibitem{Millis2} A. J. Millis, B. I. Shraiman, and R.
Mueller, Phys. Rev. Lett. {\bf 77\/}, 175 (1996).
\bibitem{REF:Holstein} T.Holstein, Ann. Phys. (N.Y.) {\bf 20} 325 (1961) 
\bibitem{REF:Friedman} 
L. Friedman and T. Holstein,  
Ann. Phys. (N.Y.) {\bf 21} 494 (1963).
\bibitem{REF:Emin}
D. Emin and T. Holstein, 
Ann. Phys. (N.Y.) {\bf 53}, 439 (1969).
\bibitem{Sheng}  
L. Sheng et al., 
Phys. Rev. Lett. {\bf 79\/}, 1710 (1997).
\bibitem{REF:Efros} 
A. Efros and B. Shklovskii, 
{\sl Electronic properties of disordered conductors\/} 
 (Springer, NY, 1984).
\bibitem{REF:LeeRama}
P. A. Lee and T. V. Ramakrishnan, Rev. Mod. Phys., {\bf 57}, 287 (1985)
\bibitem{REF:Lifshitz}
I. M. Lifshitz, 
Sov. Phys. Usp. {\bf 7\/}, 549 (1964).
\bibitem{REF:AHL}
 V. Ambegaokar, B. I. Halperin and J. S. Langer,
Phys. Rev. B {\bf 4\/}, 2612 (1971).
\bibitem{LCMOsample}  
Y. Tomioka, A. Asamitsu, and Y. Tokura, in preparation.
\bibitem{LPMOsample}  
M. Jaime, P. Lin, M. B. Salamon, and P. D. Han, Phys.
Rev. B {\bf 58}, R5901 (1998).
\bibitem{scaling}  
S. H. Chun et al., in preparation.
\bibitem{Hurd}  
C. M. Hurd, {\it The Hall Effect in Metals and Alloys}
(Plenum Press, New York, 1972).
\bibitem{REF:LeeNaga}  
N. Nagaosa and P. A. Lee, 
Phys. Rev. Lett. {\bf 64\/}, 2450 (1990).
\bibitem{REF:Wiegmann} 
Ioffe, Kalmeyer, Wiegmann
\bibitem{REF:Dagotto} 
E. M\"uller-Hartmann and E. Dagotto, 
Phys. Rev. B {\bf 54\/}, 6819 (1996).   
\bibitem{REF:LAG} Y. Lyanda-Geller, I.L. Aleiner and P.M. Goldbart,
Phys. Rev. Lett., {\bf 81} 3215 (1998).
\bibitem{footnote}   
Owing to the discreteness of the hopping, the quantal phases 
\cite{REF:LeeNaga,REF:Wiegmann,REF:Dagotto}might more
appropriately be termed Pancharatnam (rather than Berry) 
phases~\cite{REF:Pancha,REF:BonPan}.
\bibitem{Heffner} 
R. H. Heffner, J. E. Sonnier, D.E. MacLaughlin, 
G. J. Niewenhuys, G. Ehlers, F. Mezei, S-W. Cheong, J. S. Gardner and 
H. R\"oder, 
(preprint, 1999).
\bibitem{Lau} 
M. Lau and C. Dasgupta, 
Phys. Rev. B, {\bf 39\/}, 7212 (1989).
\bibitem{Furukawa} 
See, e.g., 
N. Furukawa J. Phys. Soc. Jpn {\bf 63}, 3214 (1994).
\bibitem{REF:GangOfFour}P. W. Anderson et al Phys. Rev. Lett., {\bf } (1979). 
\bibitem{Kramer} 
A. Mackinnon and B. Kramer, Phys. Rev. Lett., {\bf 47} 1546 (1981).
\bibitem{Murthy} It was also demonstrated by 
M. Kamal and G. Murthy,  Phys. Rev. Lett. 
{\bf 71} 1911 (1993). that suppression (absence) of hedgehogs 
results in a phase transition 
which characteristics (exponents) are different from those in 
the Heisenberg model. M. Kamal and G. Murthy,  Phys. Rev. Lett. 
{\bf 71} 1911 (1993). 
 \bibitem{REF:Abr} 
A. Miller and E. Abrahams, Phys. Rev. {\bf 120} 745 (1960).  
\bibitem{Louca} D. Louca et al., Phys. Rev. B. {\bf 56} R8475 (1997).
\bibitem{Billinge} S. J. L. Billinge et al., Phys. Rev. Lett., 
{\bf 77}, 715 (1996).
\bibitem{Lynn} J. Lynn, R. Ervin, J. Borching, Q. Huang, A. Santoro,
J. Peng and Z. Li, Phys. Rev. Lett., {\bf 76} 4046 (1996).
\bibitem{isotope} G. Zhao et al., Nature, {\bf 381} 676 (1996); G. Zhao et al.,
Phys. Rev. Lett., {\bf 78}, 955 (1997);
J. P. Franck et al Phys. Rev. B, {\bf 58} 5189 (1998);
A. M. Balagurov et al., Phys. Rev. B, {\bf 60}, 383 (1999).
\bibitem{spinwave} T. G. Perring et al. Phys. Rev. Lett., {\bf 77} 711 (1996).
\bibitem{REF:Jaime} M. Jaime et al.,  Phys. Rev. B, {\bf 60}, 1028 (1999). 
\bibitem{thermopower} J. Fontcuberta et al. Appl. Phys. Lett., {\bf 68}, 2288 
(1996).
\bibitem{Raman} S. Yoon et al., Phys. Rev. B, {\bf 58} 2795 (1998).
\bibitem{mweissman} R. D. Merithew et al., Phys. Rev. Lett., {\bf 84} 
3442 (2000). 
\bibitem{REF:Ziman}
See, e.g., J. M. Ziman, 
{\sl Elements of Advanced Quantum Theory\/}, 
Sec.~3.1.
\bibitem{REF:Schirmacher} W. Schirmacher, Phys. Rev. B {\bf 41} 2461 (1990).
\bibitem{RB} The essential 
product of three overlap integrals, sensitive to magnetic flux through 
loops of non-zero area, $V_{hk}V_{kj}V_{jh}$, can enter the 
transition rate also through the energy denominators, 
as we have invoked Brillouin-Wigner rather than 
Rayleigh-Schr\"odinger perturbation theory. However, the corresponding 
contributions to the probability is of the higher order in powers 
of small parameter of our perturbation expansion, $V_{ij}/(E_i-E_j)$, 
than terms features in Eq.~(\ref{EQ:MEeval}).  
\bibitem{Spivak} B. I. Shklovskii and B. Z. Spivak, in Hopping Transport in 
Solids, edited by M. Pollak and B. Shklovskii, Elsevier Science, NY (1991), 
p. 271.
\bibitem{REF:numphon} 
By two-phonon process (assisting the Hall effect) we mean one involving 
probabilities (computed in perturbation theory) characterized by 
4 carrier-phonon interaction vertices; 
the probability of a single-phonon process (leading to the 
longitudinal conductivity) involves 2 such vertices.  
\bibitem{exception}
As was discussed by Holstein, strictly speaking, 
there exists a possibility of obtaining imaginary 
denominator in two-phonon processes, in which additional $\delta$-function 
does not appear, but, for example, energies of two unoccupied
states in a triad are in resonance allowing elastic transition.
However, the possibility that two unoccupied
states will be in resonance is negligibly small 
in disordered strongly correlated system. 
\bibitem{occupation} 
It is interesting to mention that 
although in contributions involving three-stage processes 
only one phonon mode changes its population, both (i) and (ii)
have similar temperature dependence: the contribution 
involving interference of one and three-stage processes 
is proportional to both the population factor of the 
phonon mode which changes its population, and to the  population of 
the phonon mode which population is not changed (this mode 
is involved on the intermediate site). 
\bibitem{REF:Galperin}  
Y. M. Galperin et al., 
Sov. Phys. JETP {\bf 72}, 193 (1991).
\bibitem{REF:Entin}  
O. Entin-Wohlman, A. G. Aronov, Y. Levinson, Y. Imry,
Phys. Rev. Lett., {\bf 75}, 4094 (1995).
\bibitem{REF:IoLeMi} 
In the context of high temperature superconductors, 
OH current noise has been considered by 
L. Ioffe, G. Lesovik and A. J. Millis,  
Phys. Rev. Lett. {\bf 77\/}, 1584 (1996). 
The spin quantal phase acrued by hopping carriers on plaquettes 
of ions with disordered spins in  high temperature superconductors,
was also shown to affect the ordinary Hall effect~\cite{REF:Wiegmann}.
\bibitem{Goldhaber}
See, e.g., A. Goldhaber, 
Phys. Rev. Lett. {\bf 64\/}, 482 (1987).
\bibitem{neutron} M . R. Ibarra and J. M. De Teresa, in Colossal 
Magnetoresistance, Charge order and Related Phenomena, ed. C. N. Rao 
and R. Raveau, World Scientific (1998) Singapore.
\bibitem{Pokrovsky}  
I. F. Lyuksytov and V.L. Pokrovsky, Mod. Phys. Lett B, {\bf 13}, 379 (1999).
\bibitem{REF:Satpathy} Z. Popovich and S. Satpathy, Phys. Rev. Lett., 
{\bf 84} 1603 (1999). 
\bibitem{REF:Landau} L. D. Landau and E. M. Lifshitz, Quantum mechanics,
Pergamon Press, Oxford (1977). 
\end{references}
\end{document}